\documentclass[letterpaper,journal]{IEEEtran}
\usepackage{amsmath,amsfonts}
\usepackage{algorithmic}
\usepackage{algorithm}
\usepackage{array}
\usepackage[font=normalsize,labelfont=sf,textfont=sf]{subcaption}
\usepackage{textcomp}
\usepackage{stfloats}
\usepackage{url}
\usepackage{verbatim}
\usepackage{graphicx}
\usepackage{cite}
\usepackage{subcaption}
\usepackage{booktabs}
\usepackage{tabularx}
\usepackage{multirow}
\usepackage{setspace}
\usepackage{xcolor}

\usepackage{fancyhdr}
\begin{document}

\title{ARIQA-3DS: A Stereoscopic Image Quality Assessment Dataset for Realistic Augmented Reality}

%\title{ARIQA-3DS: A New and Realistic Stereoscopic Augmented Reality Image Quality Assessment Dataset}

\author{Aymen Sekhri$^{1, 2}$, Seyed Ali Amirshahi$^{2}$, Mohamed-Chaker Larabi$^{1}$~\IEEEmembership{Senior Member,~IEEE} \\ 
        \textit{$^{1}$CNRS, Université de Poitiers, XLIM, Poitiers, France} \\ 
        \textit{$^{2}$Norwegian University of Science and Technology, Gjøvik, Norway} 
        %\thanks{This work has been submitted to IEEE Transactions on Multimedia for possible publication.}
}
        % <-this % stops a space
%\thanks{This paper was produced by the IEEE Publication Technology Group. They are in Piscataway, NJ.}% <-this % stops a space
%\thanks{Manuscript received April 19, 2021; revised August 16, 2021.}

\markboth{Journal of \LaTeX\ Class Files,~Vol.~14, No.~8, August~2021}%
{Shell \MakeLowercase{\textit{et al.}}: A Sample Article Using IEEEtran.cls for IEEE Journals}


\maketitle

\begin{abstract}
As Augmented Reality (AR) technologies advance towards immersive consumer adoption, the need for rigorous Quality of Experience (QoE) assessment becomes critical. However, existing datasets often lack ecological validity, relying on monocular viewing or simplified backgrounds that fail to capture the complex perceptual interplay—termed visual confusion—between real and virtual layers. To address this gap, we present ARIQA-3DS, the first large stereoscopic AR Image Quality Assessment dataset. Comprising 1,200 AR viewports, the dataset fuses high-resolution stereoscopic omnidirectional captures of real-world scenes with diverse augmented foregrounds under controlled transparency and degradation conditions. We conducted a comprehensive subjective study with 36 participants using a video see-through head-mounted display, collecting both quality ratings and simulator-sickness indicators. Our analysis reveals that perceived quality is primarily driven by foreground degradations and modulated by transparency levels, while oculomotor and disorientation symptoms show a progressive but manageable increase during viewing. ARIQA-3DS will be publicly released to serve as a comprehensive benchmark for developing next-generation AR quality assessment models.
\end{abstract}
\begin{IEEEkeywords}
Augmented Reality, Image Quality Assessment, Stereoscopic, Subjective Experiments
\end{IEEEkeywords}

\vspace{-0.5cm}
\section{Introduction}
\vspace{-0.1cm}
\IEEEPARstart{I}{mmersive} media reshape digital experiences by engaging users’ perceptual and cognitive capacities to induce a strong sense of presence in virtual or augmented environments. The QUALINET White Paper~\cite{perkis2020qualinet} characterizes these experiences through properties such as immersivity, interactivity, exportability, believability, and plausibility. Complementary perspectives from Van Gisbergen~et~al.~\cite{van2016contextual} and the ITU~\cite{itu_g1035} emphasize experiential dimensions, including presence, perspective, proximity, and participation. Within the reality–virtuality continuum introduced by Milgram and Kishino~\cite{milgram1994taxonomy}, Virtual Reality (VR) presents fully synthetic environments, Augmented Reality (AR) superimposes augmented content onto the real world, and Mixed Reality (MR) integrates and anchors virtual elements within physical scenes. Together, these definitions highlight the growing ambition to blend real and virtual stimuli seamlessly to enrich the user experience.

Driven by this ambition, AR has advanced rapidly across navigation, education, and healthcare~\cite{vertucci2023history}. This growth is reflected in the market: in 2023, 1.4 billion users engaged with mobile AR, and the hardware sector is projected to grow from $2.4$ billion to over $9$ billion by 2027~\cite{statista_ar}. The emergence of new devices, such as Apple’s Vision Pro and headsets from Microsoft and Magic Leap, further exemplifies this momentum. To standardize these technologies, the ITU classifies AR displays into three categories: optical see-through, video see-through, and spatial systems. It also identifies key characteristics, including tracking, co-registration, rendering, and visualization, that are required to properly align and integrate augmented content with the real world~\cite{itu-rec-g1036}. As these systems evolve, understanding how users perceive the resulting mixed imagery becomes increasingly important.

Because AR experiences combine multimodal sensory cues, Quality of Experience (QoE) emerges as a central determinant of usability and acceptance. QoE reflects the degree of delight or annoyance that arises from how effectively a system meets users’ expectations of utility and enjoyment~\cite{perkis2020qualinet}. Important features such as human (e.g., visual function, binocular differences), content factors (e.g., real–virtual blending), media and coding (e.g., compression, resolution), system factors (e.g. stereoscopic rendering, depth range), and contextual (e.g., simulator sickness, age, emotional state) jointly shape perceived quality as described by~\cite{itu-rec-g1036}. The interplay among these dimensions creates a complex evaluation landscape, underscoring the need for rigorous assessment methodologies tailored to AR.

Recent studies further emphasize the technical and perceptual challenges inherent to AR Image Quality Assessment (IQA). Inaccurate spatial registration of augmented content may produce noticeable placement errors~\cite{el2022assessment}, while the fusion of virtual and real imagery can induce visual confusion, degrading the perceptual quality of both layers~\cite{visual_confusion}. Additional complexity arises from the diverse characteristics of augmented content~\cite{duan2022confusing}. Nonetheless, many AR image-quality studies rely on 2D displays or simplified 3D meshes over uniform backgrounds, limiting ecological validity and suppressing depth cues essential for realism~\cite{guo2016subjective,alexiou2017towards}. Although stereoscopic presentation restores depth, it introduces the risk of binocular visual confusion. These limitations highlight the need for datasets and evaluation methodologies that effectively capture the perceptual complexity of stereoscopic AR images.

To address these gaps, we introduce ARIQA-3DS, a new database and benchmark for stereoscopic AR image-quality assessment. High-resolution stereo omnidirectional images of real scenes are captured, and diverse augmented objects are overlaid under controlled fusion and degradation conditions. A large-scale subjective study with more than thirty participants is conducted using a video see-through head-mounted display, yielding Mean Opinion Score (MOS) for hundreds of stereo AR combinations. Building on insights from prior work~\cite{duan2022confusing, ariqa_pico, Hussain17082024, technologies12110216}, our contributions are as follows:
\begin{enumerate}
    \item We construct and publicly release ARIQA-3DS, the first stereoscopic AR image quality database, comprising stereo real-world captures paired with augmented content rendered at varying transparency levels and impairment severity.
    \item We conduct comprehensive subjective experiments following ITU guidelines~\cite{ITU-R_BT500_15,ITU_T_P910_2023}. The study collects MOS and incorporates simulator-sickness questionnaires~\cite{SSQ} to examine relationships between visual discomfort and perceived quality.
\end{enumerate}

All data-collection tools, the ARIQA-3DS database, benchmark models, and evaluation code will be released to support future research and guarantee reproducibility. In the rest of the paper, we first review related work on AR, including subjective AR image quality assessment. We then describe the proposed subjective study, including the source and test images, experimental design, viewing and scoring procedures, display setup, and subject training. Next, we detail the data analysis methodology, covering MOS and SOS analysis, quality-rating discriminability, and cybersickness evaluation. Finally, we conclude the paper and outline directions for future work.

\vspace{-0.5cm}
%-----------------------------------------------------
\section{Related Work}
\vspace{-0.1cm}
\subsection{Augmented Reality and Visual Confusion}
\vspace{-0.1cm}
This work focuses on head-mounted AR systems, which enrich users’ surroundings by overlaying objects into the real world through optical or video see-through displays. In these settings, the real and virtual layers are not perceived independently; rather, they interact through a perceptual mechanism commonly referred to as visual confusion. As shown by Duan et al.~\cite{duan2022confusing}, visual confusion fundamentally shapes how users interpret simultaneously superimposed stimuli and strongly influences AR image quality, legibility, and comfort. Factors such as cluttered backgrounds, limited contrast, and inconsistent depth cues can make the augmented content more difficult to perceive clearly within the real environment~\cite{cooper2023perceptualAR}. These perceptual interactions differ from those in conventional 2D IQA, underscoring the need for AR-specific models that account for the perception of both real and augmented content.

A key aspect of visual confusion is that it manifests in two principal forms: monocular and binocular. Monocular visual confusion occurs when two views—typically a real-world scene and augmented content—are superimposed within a single eye. According to Duan et al. \cite{duan2022confusing}, this leads to monocular rivalry, a state where overlapping patterns compete for perceptual dominance, ultimately impacting the user's QoE. This is often observed when semi-transparent overlays or low-contrast virtual elements are superimposed on textured, high-frequency real environments. In contrast, binocular visual confusion occurs when two different views are presented to each eye, respectively. Such confusion, stemming from stereoscopic disparity errors, incorrect occlusion handling, non-uniform transparency, or misaligned rendering, may exceed the limits of binocular fusion. This can trigger binocular rivalry, alternating perceptual dominance, depth distortions, and visual discomfort \cite{fezza_larabi}. Together, these monocular and binocular forms of confusion govern the perceptual stability of AR content, the detectability and interpretability of augmented objects, and the overall QoE. Their influence has motivated recent AR-IQA research to incorporate real-virtual interaction effects and confusion-related mechanisms into dataset design and objective metric development \cite{duan2022confusing}.
\vspace{-0.5cm}
\subsection{Subjective AR Image Quality Assessment}
\vspace{-0.1cm}
Although Subjective Quality Assessment (SQA) has been extensively studied in VR, 360° media, and stereoscopic 3D imaging, its application to AR remains comparatively underexplored. Most existing works evaluate geometric degradations of isolated digital objects, such as point clouds and 3D meshes rendered on blank or uniformly textured backgrounds. Guo et al.~\cite{guo2016subjective} examined the visual quality of distorted meshes through controlled rotation sequences, while Alexiou et al.~\cite{alexiou2017towards} extended this line of research to point-cloud data by studying the perceptual impact of geometric and texture impairments. More recently, Ak et al.~\cite{ak2024basics} introduced BASICS, one of the largest SQA datasets for static point clouds in compression evaluation, though the content still lacks realistic AR integration and does not address perceptual conflicts between virtual and real layers.

A major shift occurred with the work of Duan et al.~\cite{duan2022confusing}, who moved AR-IQA research beyond classical geometric degradations to focus on visual confusion, a perceptual artifact inherent to AR’s superimposed viewing paradigm. Unlike earlier studies that considered isolated objects against blank or minimally textured backgrounds, their work conceptualizes AR as the perceptual fusion of a foreground augmented layer and a background see-through layer. To study this interaction systematically, they first introduced the ConFusing Image Quality Assessment (CFIQA) dataset, aimed at understanding how humans perceive superimposed pairs of images. Building on these insights, Duan et al. later constructed the ARIQA dataset to emulate realistic AR conditions more realistically. ARIQA contains 20 augmented images, 20 real background images, and 560 distorted AR stimuli generated by systematically mixing AR and background layers across four confusion levels and multiple distortions, including JPEG compression, image scaling, and contrast adjustment. Unlike CFIQA, where both layers share identical fields of view, ARIQA approximates real AR display geometry by assigning a larger FOV to the background layer, consistent with see-through optics. To preserve experimental control, the subjective study (23 observers) was carried out in a VR environment, enabling real–virtual fusion conditions. Although the ARIQA dataset emulates AR quality more realistically, several limitations remain. First, the dataset is limited to monocular viewing; subjects are presented with the same view for both eyes, which ignores a critical aspect of realism: depth perception. In addition, the ARIQA dataset impaired only the augmented objects while assuming the background remains pristine, which is not always the case in AR scenarios where the real-world background is also prone to various impairments. Finally, content diversity is a key factor for simulating various AR scenarios, which remains limited in this dataset.

To better approximate real AR usage, the ARIQA-PICO dataset~\cite{ariqa_pico} was recently introduced. Unlike the VR-simulated ARIQA stimuli, ARIQA-PICO captures genuine mixed-reality views using the PICO4 optical see-through headset. The dataset comprises 450 AR scenes overlaying 15 virtual images from three semantic categories (natural, web, and cartoon) onto 10 real scenes, each rendered at three transparency levels $\lambda \in \{0.25, 0.5, 0.75\}$. A subjective study with 20 observers yielded 900 MOS values, revealing three dominant perceptual regime:, unilateral suppression, acceptable confusion, and intense confusion. Notably, medium transparency ($\lambda = 0.5$) produced the strongest interference, where both real and virtual components degrade simultaneously in perceived quality. Table~\ref{tab:dataset_overview} summarizes a comparison between ARIQA~\cite{duan2022confusing}, ARIQA-PICO~\cite{ariqa_pico}, and our proposed dataset, ARIQA-3DS.

\begin{figure*}[!t]
    \centering
    \begin{subfigure}[t]{0.3\textwidth}
        \centering
        \includegraphics[width=0.9\linewidth]{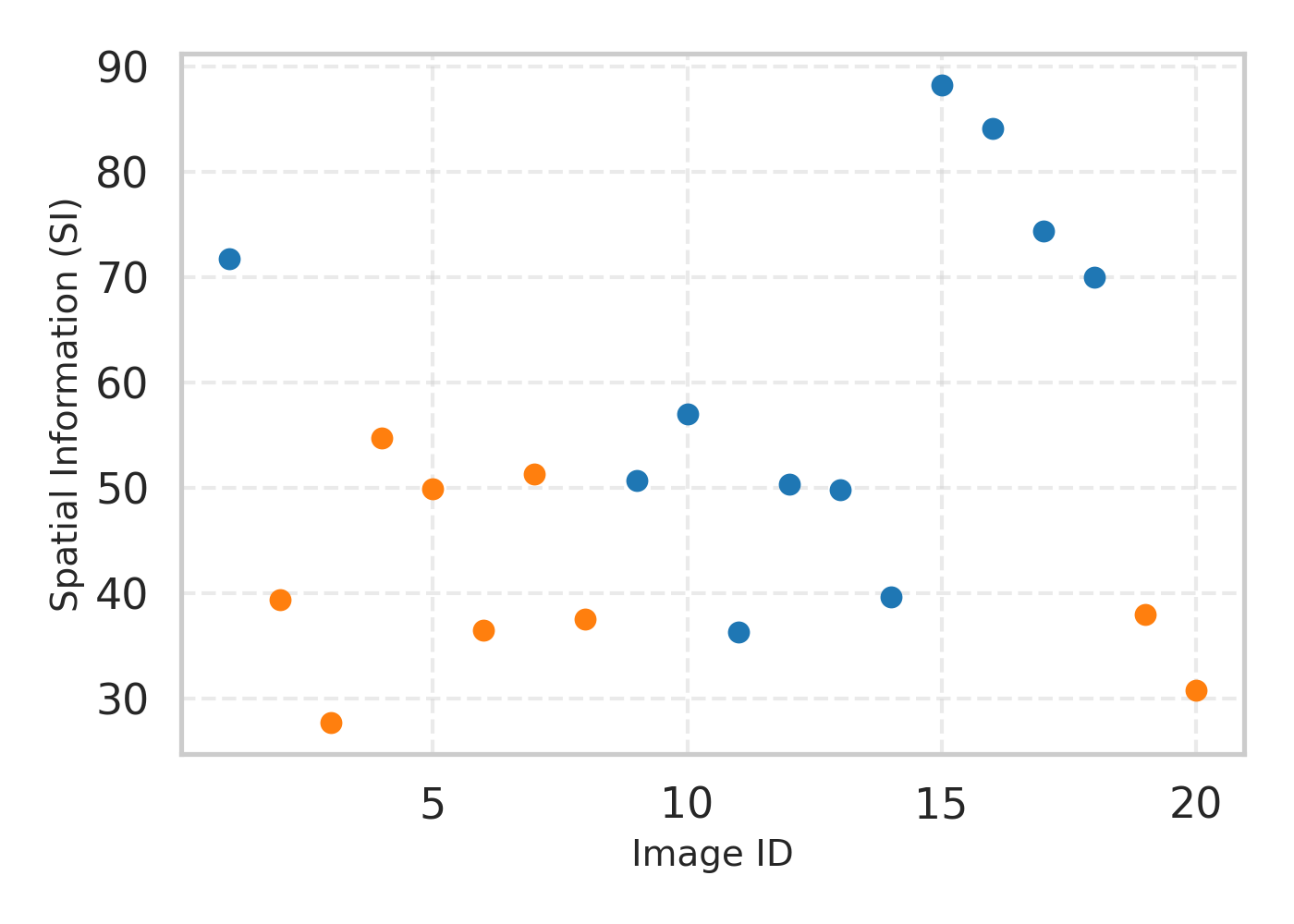}
        \caption{\small SI}
        % \caption*{\normalsize Spatial Information (SI)}
        \label{fig:si_bg}
    \end{subfigure}
    \hfill
    \begin{subfigure}[t]{0.3\textwidth}
        \centering
        \includegraphics[width=0.9\linewidth]{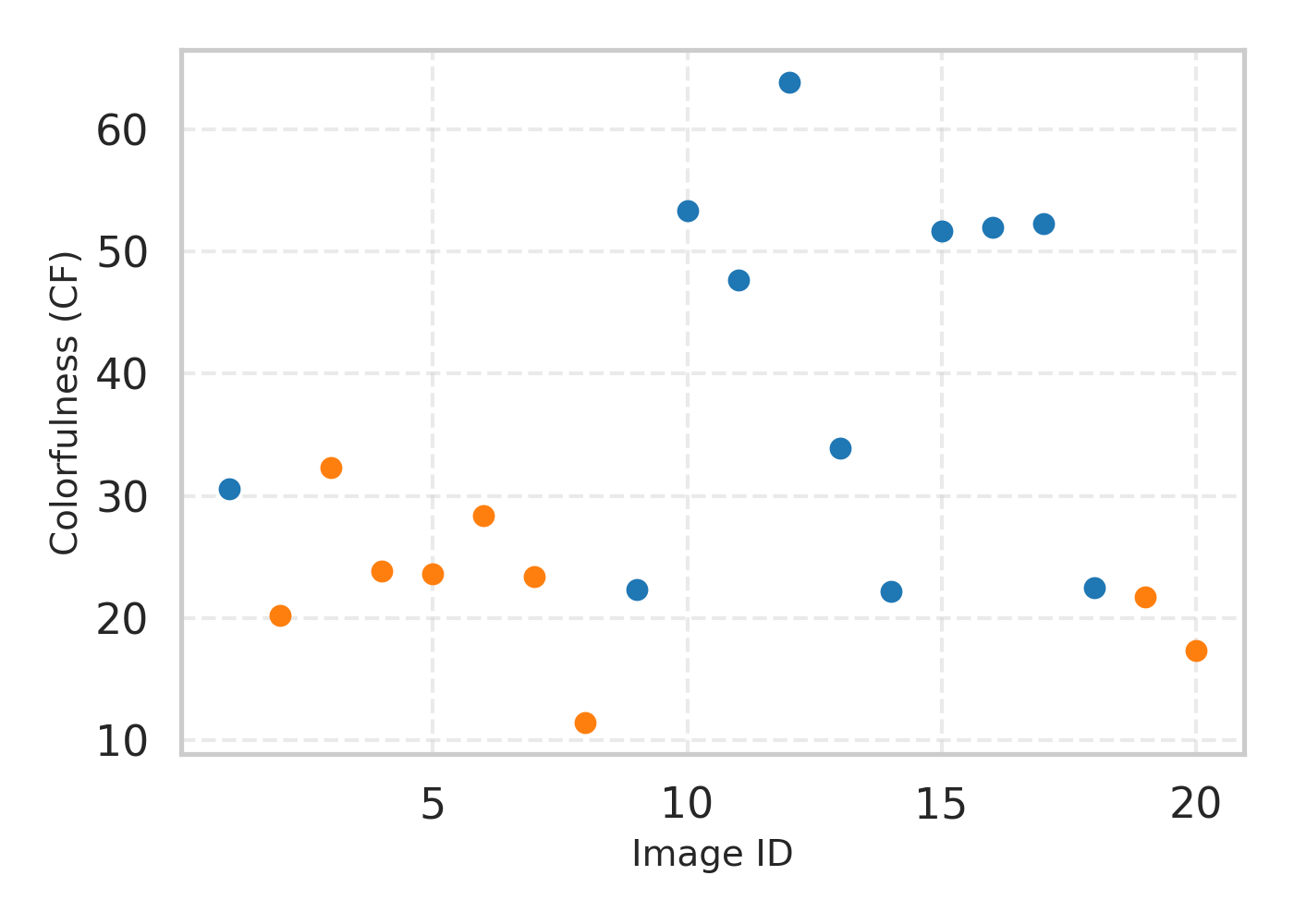}
        \caption{\small CF}
        % \caption*{\normalsize Colorfulness (CF)}
        \label{fig:cf_bg}
    \end{subfigure}
    \hfill
    \begin{subfigure}[t]{0.3\textwidth}
        \centering
        \includegraphics[width=0.9\linewidth]{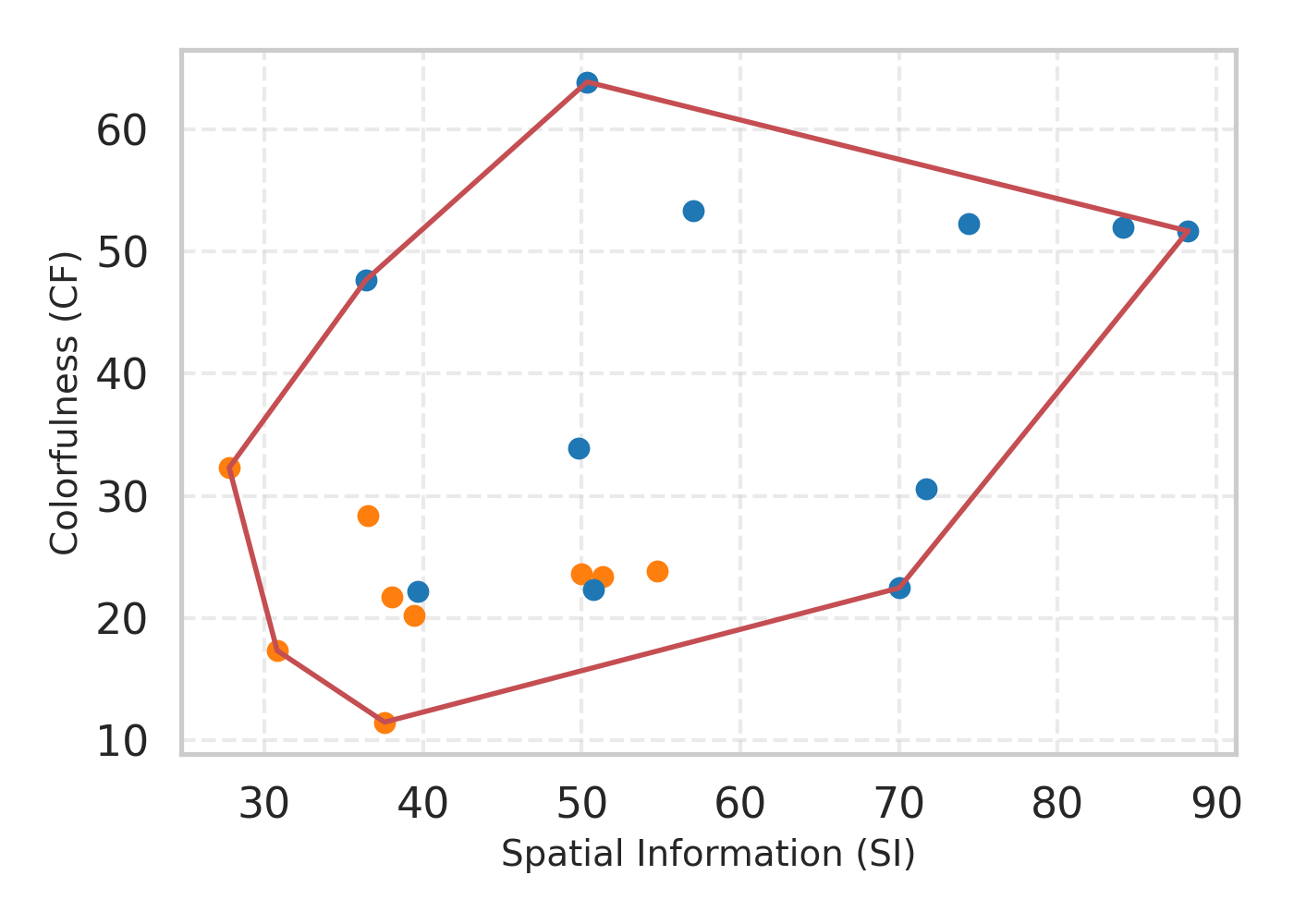}
        \caption{\small SI $\times$ CF}
        % \caption*{\normalsize Joint SI–CF distribution}
        \label{fig:si_cf_bg}
    \end{subfigure}
    \caption{Plots of Spatial Information (SI) and Colorfulness (CF) for the twenty stereoscopic $360^{\circ}$ background images (orange: indoor scenes, blue: outdoor scenes). (a) SI index shows a wide range of structural complexity; (b) CF metric varies from low to high color saturation; (c) the joint distribution indicates that the selected images span a broad region of the SI $\times$ CF space.}
    \label{fig:si_cf_plots_bg}
\end{figure*}

\begin{figure*}[!t]
    \centering
    \footnotesize
    \begin{subfigure}[t]{0.3\textwidth}
        \centering
        \includegraphics[width=0.9\linewidth]{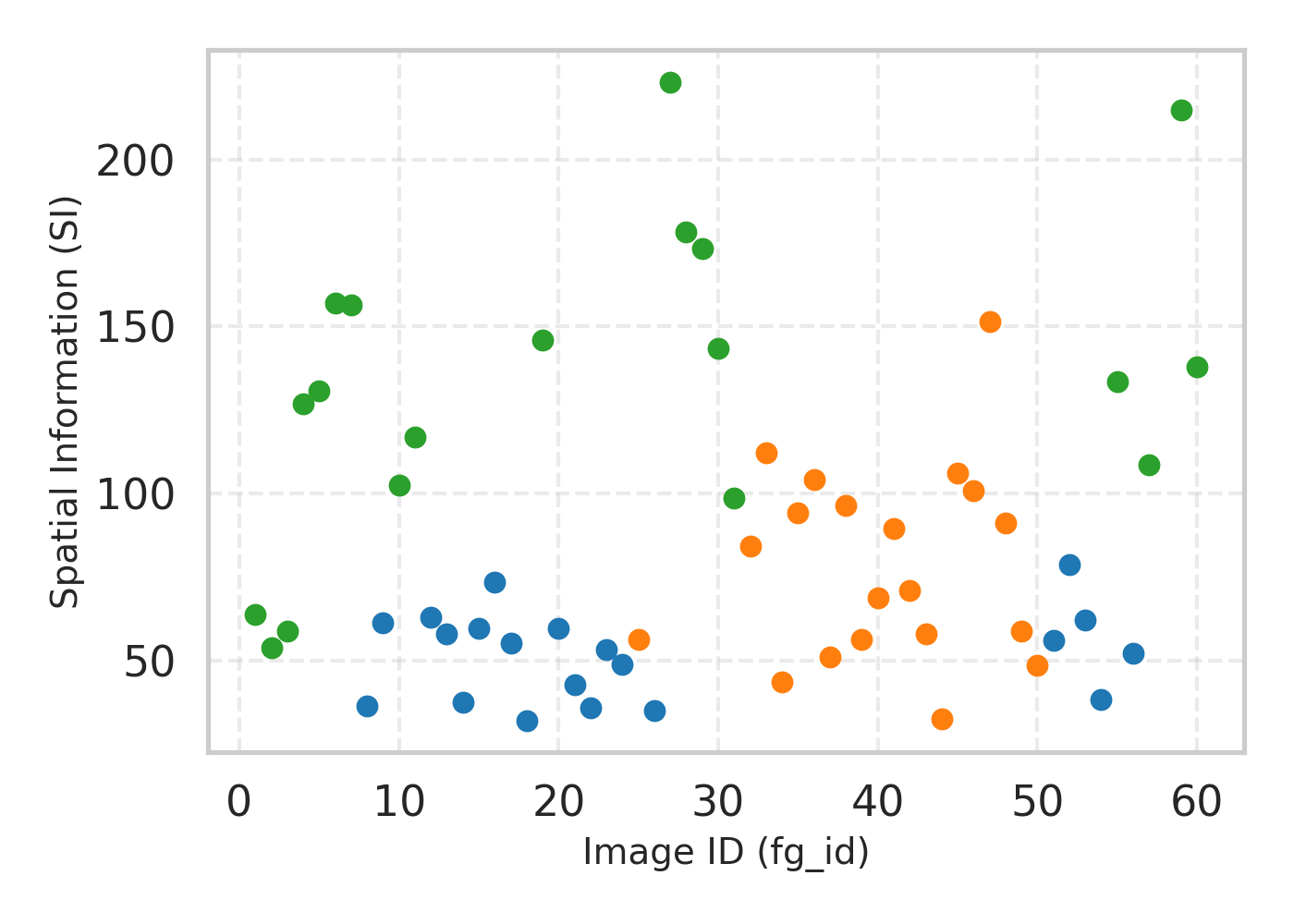}
        \caption{\footnotesize SI}
        \label{fig:si_fg}
    \end{subfigure}
    \hfill
    \begin{subfigure}[t]{0.3\textwidth}
        \centering
        \includegraphics[width=0.9\linewidth]{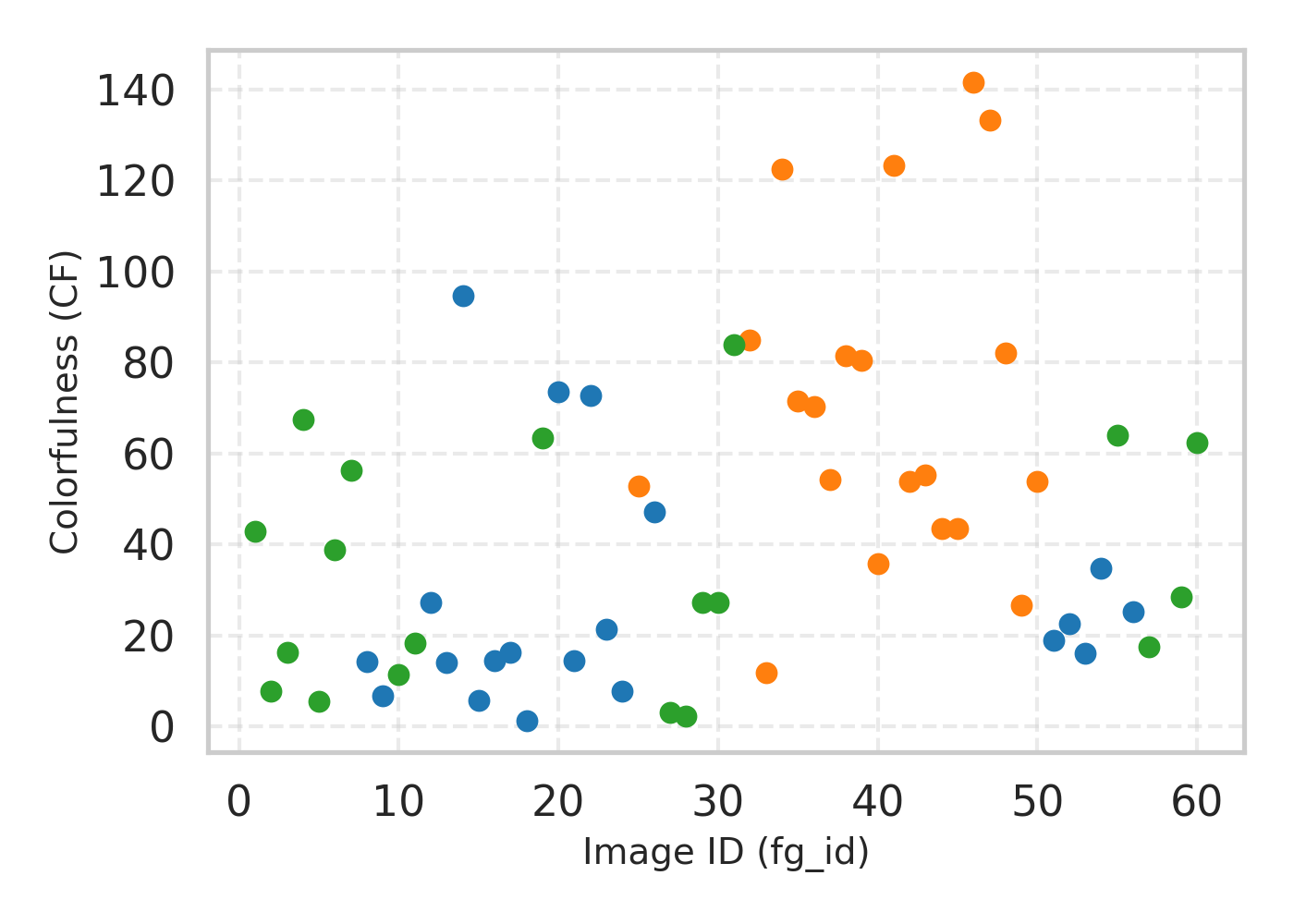}
        \caption{\small CF}
        \label{fig:cf_fg}
    \end{subfigure}
    \hfill
    \begin{subfigure}[t]{0.3\textwidth}
        \centering
        \includegraphics[width=0.9\linewidth]{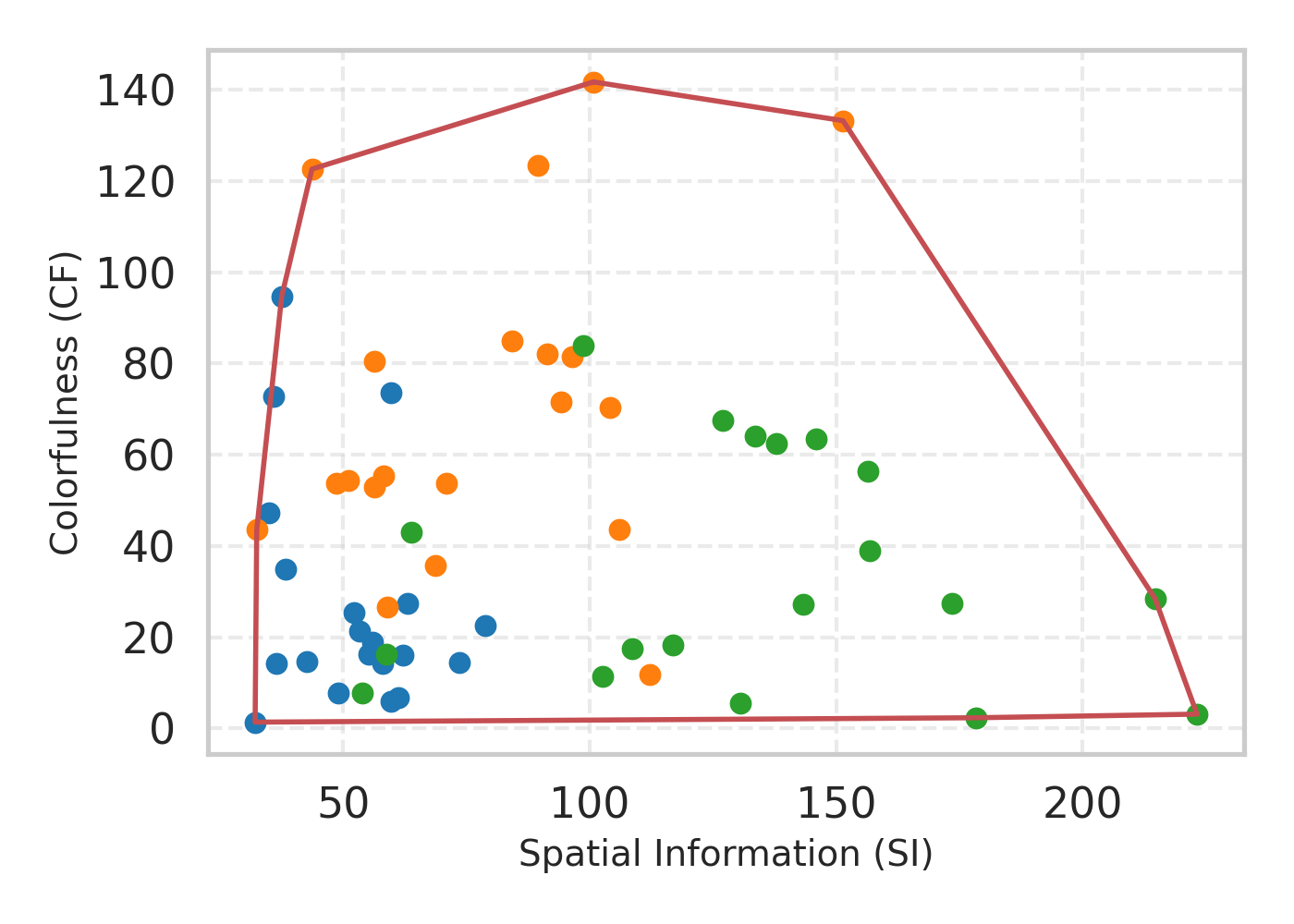}
        \caption{\small SI $\times$ CF}
        \label{fig:si_cf_fg}
    \end{subfigure}
    \caption{Plots of Spatial Information (SI) and Colorfulness (CF) for the sixty foreground objects (blue: Graphical, orange: Natural, green: Screenshots). The three semantic categories occupy distinct regions in the SI $\times$ CF feature space, highlighting the diversity of structural textures and chromatic properties represented in the dataset.}
    
    \label{fig:si_cf_plots_fg}
\end{figure*}

\vspace{-0.5cm}
\section{Subjective Study}
\vspace{-0.1cm}
This section describes the construction of the proposed stereoscopic Augmented Reality Image Quality Assessment (ARIQA-3DS) dataset, the degradations applied to simulate practical AR rendering effects, and the subjective evaluation protocol used to collect perceptual quality scores. The following subsections describe the acquisition setup, detail the distortion models applied to real--virtual composites, and outline the experimental procedure, demonstrating how these components collectively ensure a controlled evaluation of stereoscopic AR quality.
\vspace{-0.5cm}
\subsection{Source Images}
Constructing realistic AR scenes requires both a credible background layer and a representative set of foreground objects. To this end, we collected stereoscopic $360^{\circ}$ background images together with three distinct classes of augmented objects, and analyzed them using perceptually motivated descriptors to verify their diversity and suitability for AR-quality evaluation.

\begin{figure*}[t]
    \centering
    \includegraphics[width=\textwidth]{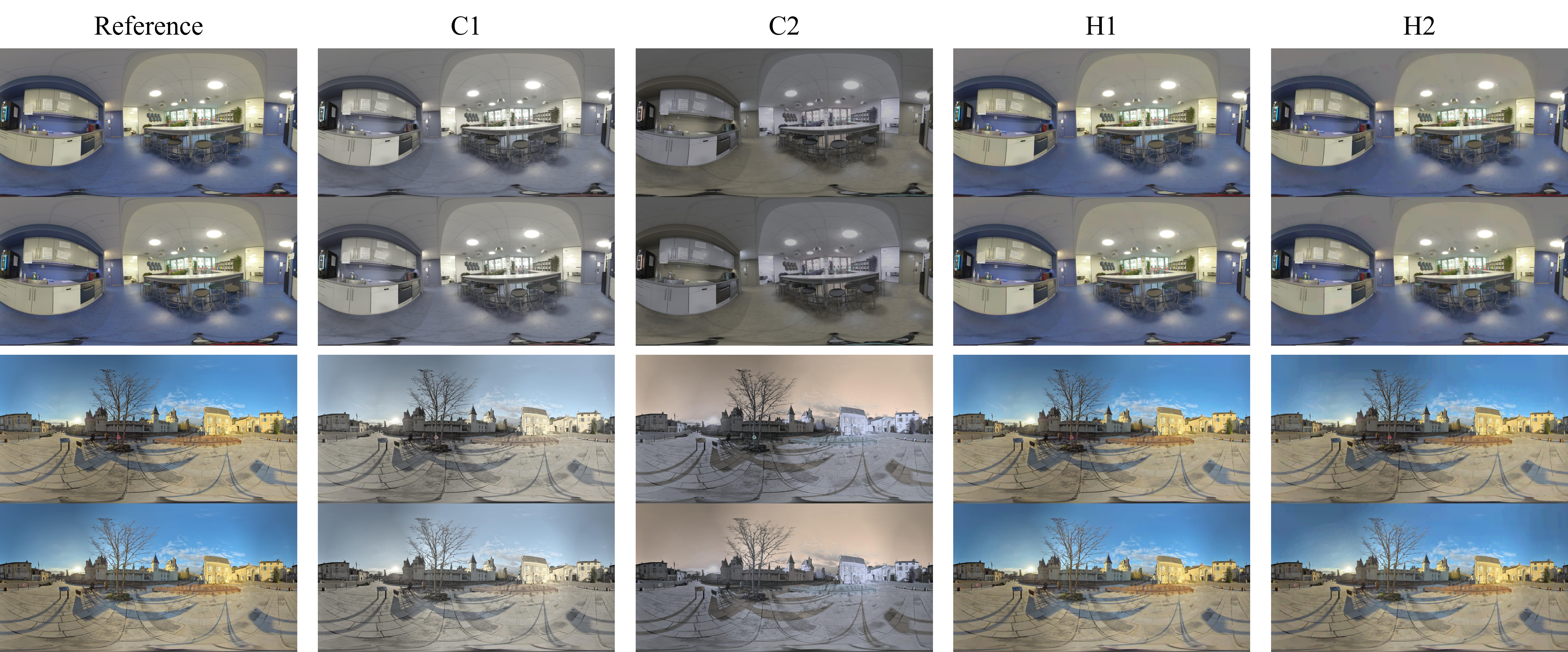}
    \caption{Representative stereoscopic $360^{\circ}$ background images captured with Insta360 Pro~1 and Insta360 Pro~2 cameras in indoor and outdoor scenes from Poitiers (France), and Gjøvik (Norway). The reference images are shown alongside distorted versions generated using two color saturation levels (C1, C2) and two HEVC compression levels (H1, H2).\vspace{-0.5cm}}
    \label{fig:BGSamples}
\end{figure*}

\begin{figure}[t]
    \centering
    \includegraphics[width=0.45\textwidth]{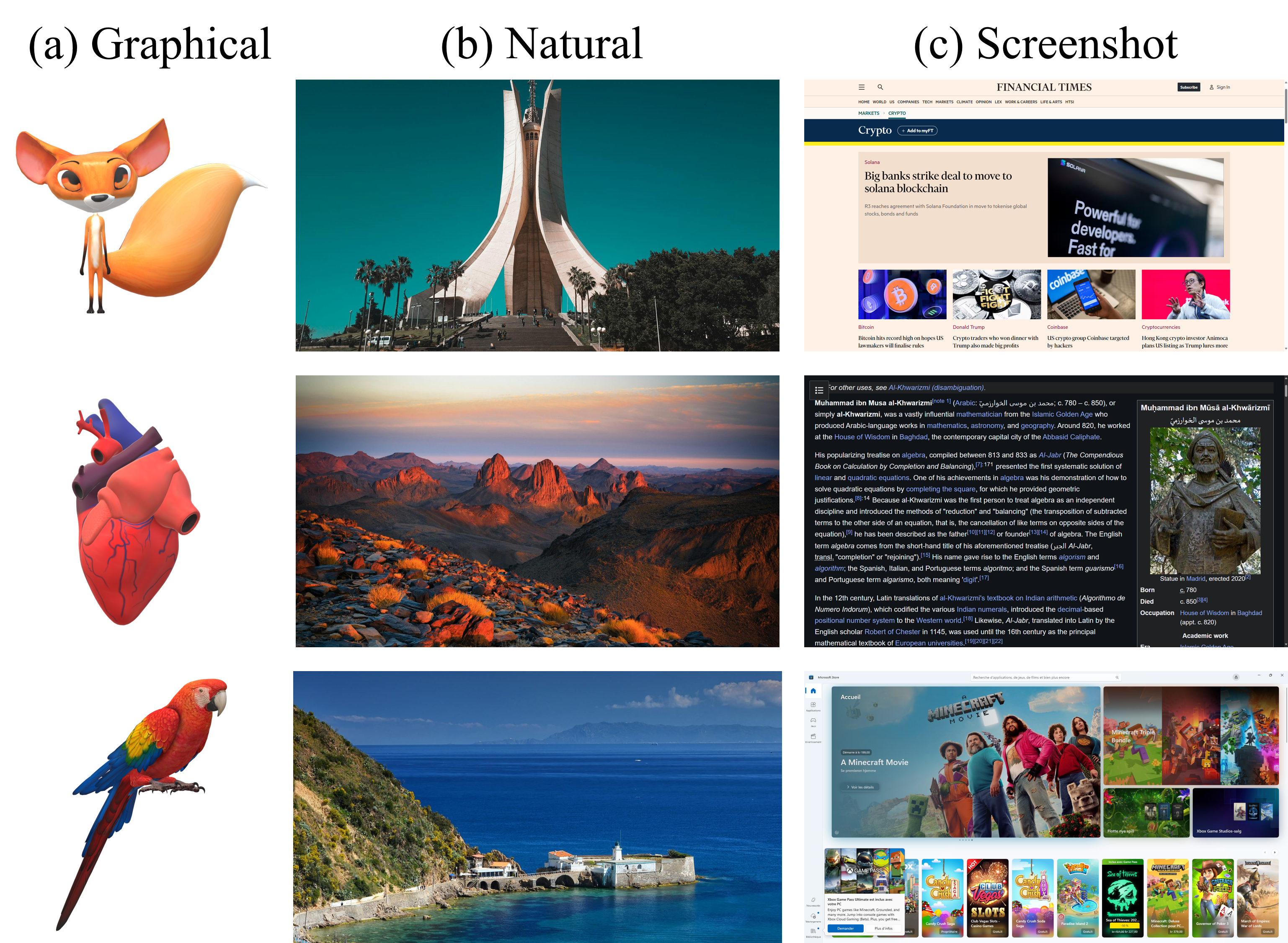}
    \caption{Virtual foreground objects included in ARIQA-3DS, spanning three semantic categories: 
    Graphical (3D Paint Microsoft and Sketchfab\protect\footnotemark[1]), 
    Natural (Pexels\protect\footnotemark[2]), 
    and Screenshot.\vspace{-0.5cm}}
    \label{fig:FGSamples}
\end{figure}

\footnotetext[1]{Sketchfab: \url{https://www.sketchfab.com}}
\footnotetext[2]{Pexels: \url{https://www.pexels.com}}

\paragraph{Stereoscopic Backgrounds}
We captured twenty uncompressed stereoscopic $360^{\circ}$ background images at a resolution of $7680\times3840$ using Insta360~Pro~1 and Insta360~Pro~2 cameras. Recordings are conducted across indoor and outdoor locations in Poitiers (France) and Gj{\o}vik (Norway), covering both open landscapes and cluttered indoor spaces. Unlike datasets that emphasize visually striking landmarks, we deliberately selected everyday scenes to promote natural variation in structural details and colorfulness. Representative samples are shown in Figure~\ref{fig:BGSamples}. To characterize the content of each panorama, we computed two objective descriptors widely used in image-quality assessment, the Spatial Information (SI) index~\cite{ITU_T_P910_2023}, which measures structural complexity and edge activity, and the Colorfulness (CF) metric of Hasler and Süsstrunk \cite{hasler2003measuring}, which quantifies chromatic richness. Figures~\ref{fig:si_bg} and \ref{fig:cf_bg} report SI and CF values for all twenty backgrounds, revealing substantial variation from low-structure, low-colorfulness indoor environments to highly textured and vivid outdoor scenes. The joint SI--CF distribution in Figure~\ref{fig:si_cf_bg} confirms that the selected panoramas span a broad perceptual range, ensuring that subsequent AR composites are evaluated under diverse structural and chromatic conditions.

\paragraph{Augmented Contents}
To generate realistic AR composites, we collected sixty foregrounds spanning three semantic categories: Graphical objects from the Microsoft 3D Objects repository and Sketchfab\protect\footnotemark[1], Natural objects sourced from the Pexels\protect\footnotemark[2] photo library, and Screenshot elements captured from laptop interfaces. These categories reflect common AR use cases, including synthetic 3D models, natural photographic content, and planar UI components. Representative examples are shown in Figure~\ref{fig:FGSamples}. To evaluate the diversity of the overlays, we computed the SI and CF measures as previously described. As shown in Figs.~\ref{fig:si_fg} and \ref{fig:cf_fg}, the objects represent a broad range of structural complexity and chromatic richness. The joint SI--CF distribution in Figure~\ref{fig:si_cf_fg} shows that the three foreground categories occupy distinct but partially overlapping regions in the feature space, confirming the diversity in terms of structural and chromatic characteristics is guaranteed. It is noteworthy that since the 360-degree background is configured for stereoscopic over-under rendering, the augmented content is dynamically rendered in stereo by the Unity engine to maintain binocular depth consistency.

\vspace{-0.5cm}
\subsection{Test Images}
\vspace{-0.1cm}
To approximate the types of artifacts arising in AR systems, we systematically degrade the background panoramas and foreground objects, then fuse them at two predefined transparency levels. These synthesized AR viewports form the full stimulus set for the subjective evaluation. % and provide the data foundation for objective metric benchmarking.

\paragraph{Background distortions}
Two representative degradations are applied to the real stereoscopic backgrounds. First, HEVC compression is introduced using FFmpeg’s \texttt{libx265} encoder with quantization parameters $\mathrm{QP}=40$ and $50$, where larger QP values yield stronger compression and lower perceptual quality. Second, color saturation is altered in the HSV domain to emulate chromatic variations arising from illumination changes or camera post-processing. For this, two desaturation and two oversaturation levels are produced while keeping hue and luminance fixed, ensuring that only color vividness is modified.

\paragraph{Foreground distortions}
For the augmented content, we apply distortions representative of rendering artifacts and resource-constrained AR systems. Pixelation is introduced at two severity levels $s \in \{0.1, 0.5\}$, by downscaling each object using the distortion-driven factor $z = 0.95 - s^{0.6}$~\cite{kadid10k} and subsequently upsampling with nearest-neighbour interpolation, producing blocky structures characteristic of low-resolution rendering. Motion blur is simulated by convolving each object with a directional blur kernel whose angle is randomly assigned, mimicking rapid motion or unstable rendering pipelines. These distortions capture common visual artifacts observed in AR systems operating under latency or computational constraints.

\paragraph{Alpha blending}
\label{sec:alpha_blend}
The distorted foregrounds and backgrounds are combined to form the final AR composites using
\begin{equation}
    I_M = \alpha \circ D_{\mathrm{A}}(I_A) + (1 - \alpha) \circ D_{\mathrm{B}}(I_B),
    \label{eq:alpha_blending}
\end{equation}
where $D_{\mathrm{A}}$ and $D_{\mathrm{B}}$ denote the distortion operators applied to the virtual and background layers, respectively, and $\alpha \in \{0.50, 0.75\}$ specifies the transparency level. Lower values of $\alpha$ increase the dominance of the background, whereas higher values enhance the visibility of the foreground. Manipulating $\alpha$ enables a controlled exploration of visual confusion, a perceptual phenomenon in which competing features from real and virtual layers interfere with clarity and interpretation~\cite{duan2022confusing,Hussain17082024,ariqa_pico}. The selected transparency levels approximate common conditions in optical and video see-through AR systems, where appropriate layer fusion is critical for stable perception.

\paragraph{Dataset construction}
The final dataset is constructed from 60 reference augmented contents and 20 reference backgrounds. Each augmented content is processed under two distortion types at two severity levels, yielding $60 \times 2 \times 2 = 240$ distorted variants plus the original versions, totaling 300 augmented content instances. Similarly, each background undergoes two distortion types at two severity levels, producing $20 \times 2 \times 2 = 80$ distorted variants in addition to the 20 originals, resulting in 100 background instances. To construct the AR viewports, the 60 augmented contents are randomly partitioned into 20 groups of three objects. Each triplet is paired with a single background, and the three augmented contents are placed at distinct viewport positions (front, left, and right), as shown in Figure~\ref{fig:viewportIllustration}. For every foreground-background combination, all distortion configurations are generated and blended using two transparency levels, $\alpha$. This procedure produces 600 blended viewports per session. Repeating the random grouping process produces more AR scenarios, resulting in a total of 1,200 unique AR viewports in the ARIQA-3DS dataset. 

Table~\ref{tab:dataset_overview} provides a comparative summary of major AR image-quality assessment datasets, highlighting the substantial expansion in scope achieved by ARIQA-3DS. Unlike ARIQA and ARIQA-PICO, which rely on monoscopic rendering and offer limited variation in augmented content and background complexity, ARIQA-3DS introduces a significantly broader range of foreground objects, diverse real-world stereoscopic backgrounds, and controlled distortions applied to both layers. The use of stereoscopic rendering further supplies depth cues essential for realistic AR perception. Combined with a larger number of subjects and a wider set of blending configurations, ARIQA-3DS offers a more comprehensive and ecologically valid foundation for studying AR quality and visual confusion.

\begin{table*}[t]
    \centering
    \caption{Summary of major AR image-quality assessment datasets and their characteristics.}
    \label{tab:dataset_overview}

    \setlength{\tabcolsep}{3pt}
    \renewcommand{\arraystretch}{1.1}

    \begin{tabularx}{\textwidth}{l c c c c c c c c c}
        \toprule
        \textbf{Dataset} &
        \textbf{Year} &
        \textbf{HMD} &
        \textbf{\# Viewports} &
        \textbf{\# FGs} &
        \textbf{\# BGs} &
        \textbf{\# Distortions} &
        \textbf{\# Alpha} &
        \textbf{Stereo} &
        \textbf{\# Subjects} \\
        \midrule

        ARIQA~\cite{duan2022confusing} &
        2022 &
        HTC Vive Pro Eye &
        560 &
        20 &
        20 &
        3 &
        4 &
        No &
        23 \\

        ARIQA-PICO~\cite{ariqa_pico} &
        2024 &
        PICO4 &
        450 &
        15 &
        10 &
        0 &
        3 &
        No &
        20 \\

        ARIQA-3DS (Ours) &
        2025 &
        Varjo VR-3 &
        1200 &
        60 &
        20 &
        4 &
        2 &
        Yes &
        36 \\
        
        \bottomrule
    \end{tabularx}
\end{table*}

\vspace{-0.5cm}
\subsection{Subjective Testing Design}

Our subjective evaluation follows ITU-R Recommendation BT.500~\cite{ITU-R_BT500_15} and employs a Single Stimulus (SS) methodology, where each stimulus is rated independently without an explicit reference. This protocol aligns closely with natural AR usage, where observers view a single fused scene rather than side-by-side comparisons, and it supports the efficient assessment of large stimulus sets under consistent viewing conditions. Direct testing on AR devices often introduces variability stemming from ambient illumination, head-motion differences, and limited control over background environments~\cite{duan2022confusing,ariqa_pico}. To mitigate these issues and ensure strict experimental reproducibility, all AR composites are simulated within a controlled VR environment using a Varjo~VR-3 headset.

\begin{figure}[t]
    \centering
    \includegraphics[width=0.4\textwidth]{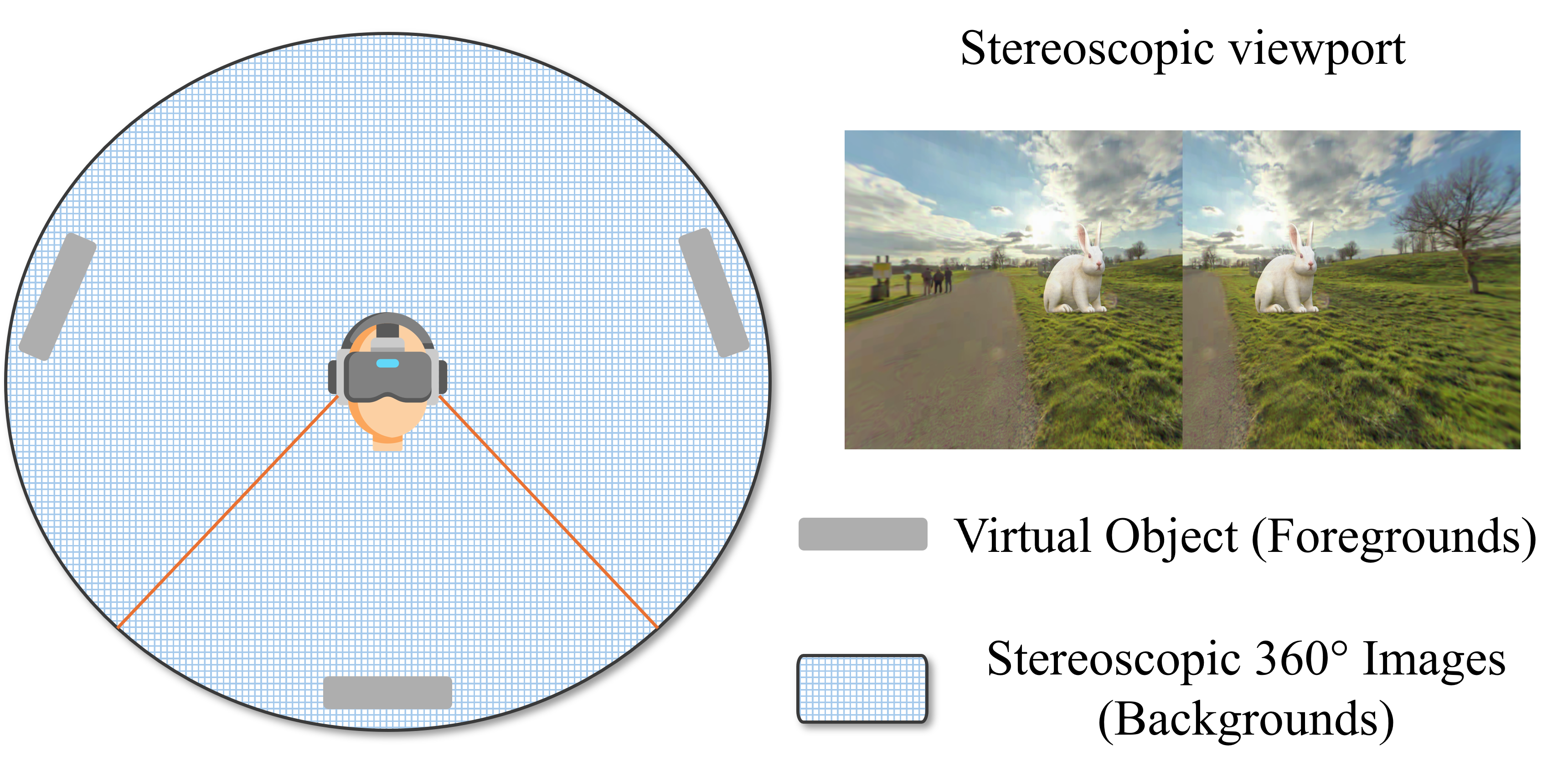}
    \caption{Illustration of AR simulation within the VR environment. Three foregrounds superimposed on a stereoscopic background. The subject focuses on a perceptual viewport, automatically shifting to the next Foreground after each rating.\vspace{-0.5cm}}
    \label{fig:viewportIllustration}
\end{figure}

Figure~\ref{fig:viewportIllustration} illustrates the simulated AR display in our study. Each stereoscopic background is rendered on the Varjo~VR-3 headset with the central display disabled, while three augmented contents are superimposed at fixed positions so that only one is visible within the user's field of view. Observers are instructed to evaluate only the object directly in front of them. After each rating, the viewport automatically shifts to the next augmented content, eliminating the need for physical head movement. This constrained viewing setup minimizes variations in head pose and viewing distance, ensuring that subjective scores capture perceptual differences arising from AR distortions rather than user behavior~\cite{perkis2020qualinet}.

\vspace{-0.5cm}
\subsection{Viewing and Scoring}
\begin{figure*}[t]
    \centering
    \includegraphics[width=0.85\textwidth]{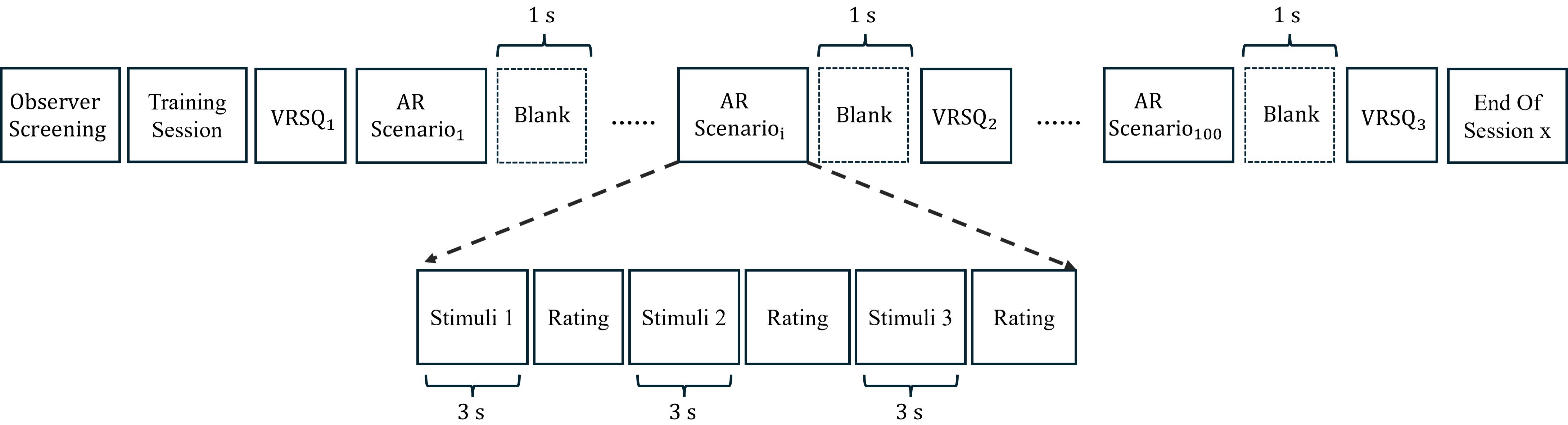}
    \caption{Overview of the subjective experiment protocol.  Each participant begins with observer screening and an initial VRSQ ($\mathrm{VRSQ}_1$), evaluates 100 AR scenarios and completes intermediate ($\mathrm{VRSQ}_2$) and final ($\mathrm{VRSQ}_3$) questionnaires to track discomfort during VR usage.  Each scenario contains three foreground objects followed by rating tasks.\vspace{-0.5cm}}
    \label{fig:subjective_plan}
\end{figure*}

\vspace{-0.1cm}
Before data collection, each observer’s visual acuity and Color vision is verified to ensure the reliability of the subjective scores. All observers provided written informed consent prior to participation. Observers are then briefed on the task and asked to complete a short training session in which they view sample AR stimuli representative of the distortions used in the study. These training stimuli are excluded from the final analysis and serve only to familiarize participants with the test environment and the scoring interface. To monitor cybersickness throughout the experiment, participants complete the Virtual Reality Sickness Questionnaire (VRSQ) immediately before the test, at the mid-point, and after completion. The VRSQ consists of nine symptoms grouped into oculomotor ($O$), disorientation ($D$), and total score ($TS$) ~\cite{SSQ}. The overall experimental protocol is summarized in Figure~\ref{fig:subjective_plan}.

\begin{figure}[t]
    \centering
    \begin{subfigure}[t]{0.45\textwidth}
        \centering
        \includegraphics[width=\linewidth]{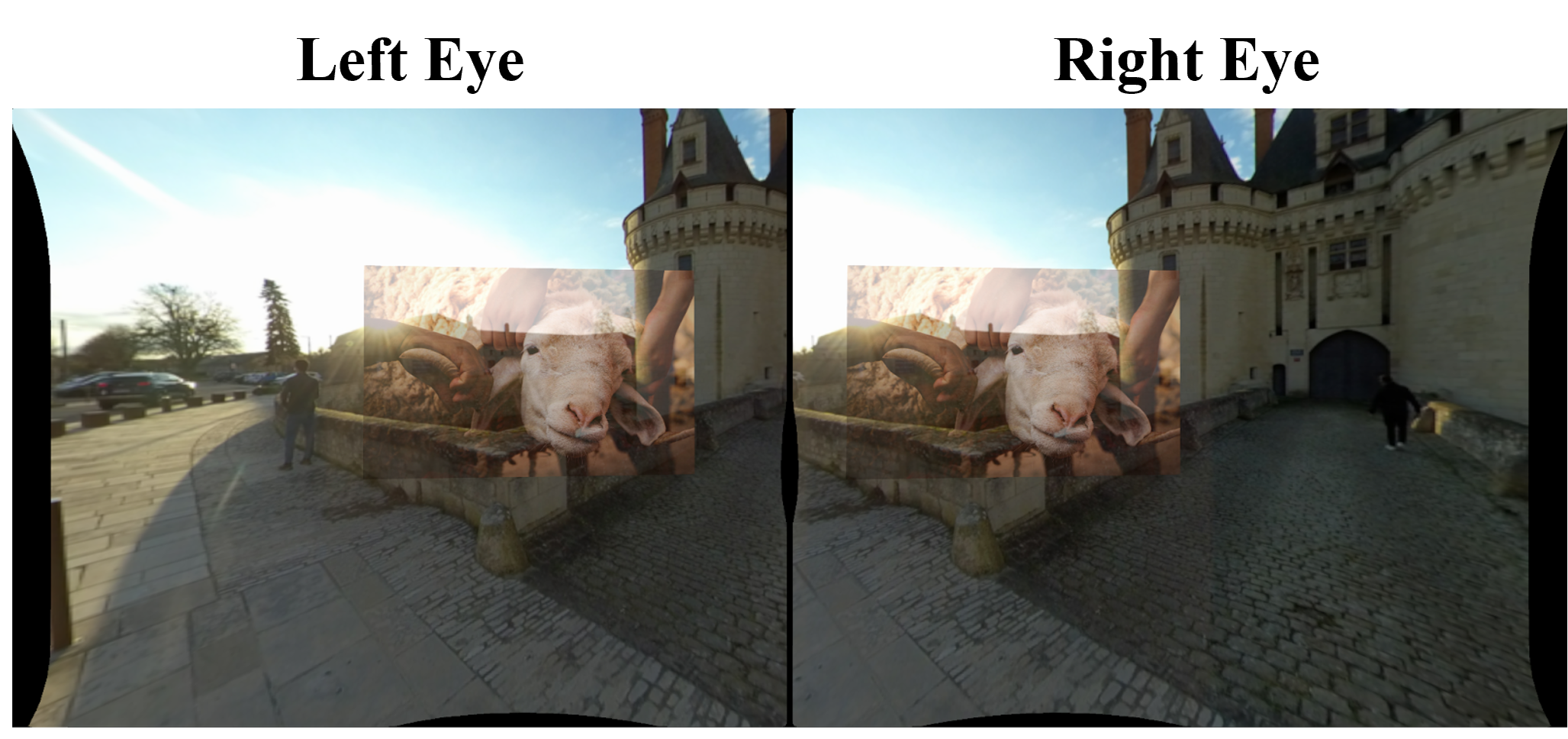}
        \caption{\footnotesize Observing}
        \label{fig:observe_phase}
    \end{subfigure}
    \hfill
    \begin{subfigure}[t]{0.45\textwidth}
        \centering
        \includegraphics[width=\linewidth]{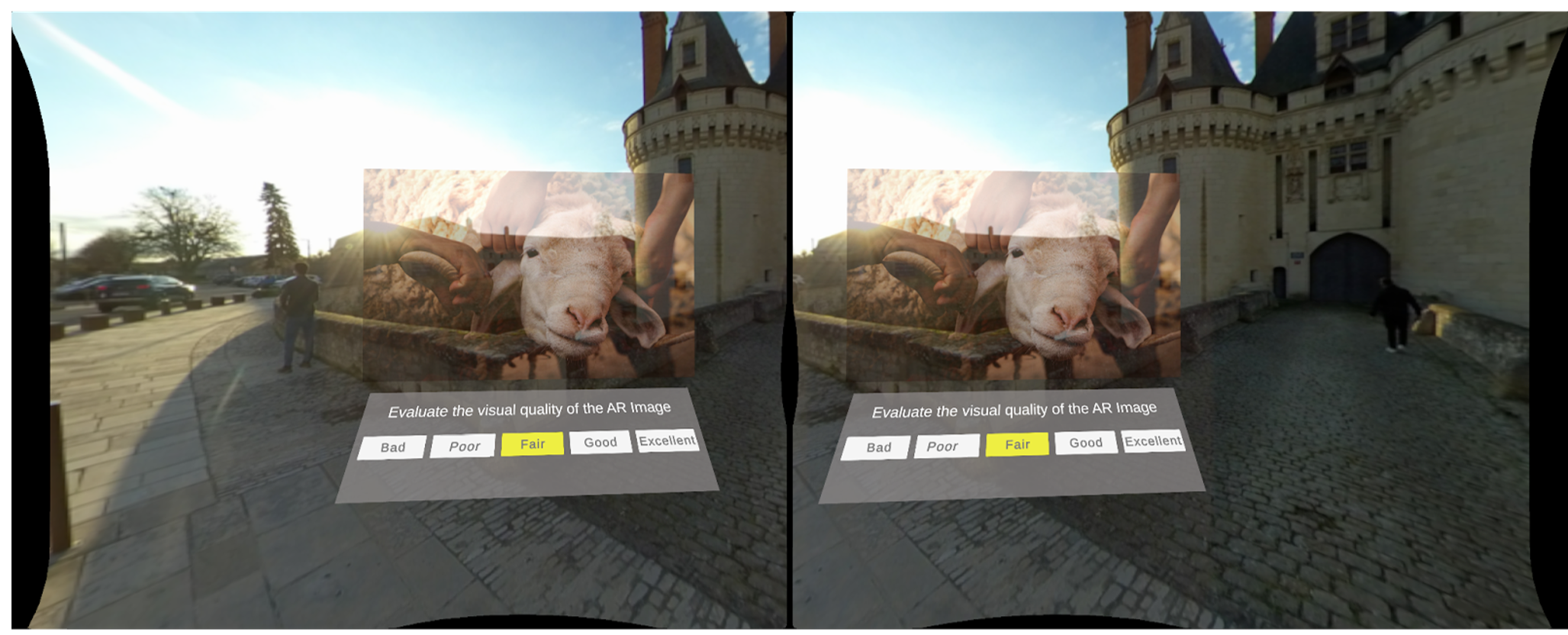}
        \caption{\footnotesize Rating}
        \label{fig:rating_phase}
    \end{subfigure}
    \caption{Stereoscopic views (left and right eye) captured from the head-mounted display during the subjective experiment. (a) Observing phase, where participants view the AR stimulus in the $360^{\circ}$ environment. (b) Rating phase, where participants evaluate visual quality using the assessment interface.\vspace{-0.5cm}}
    \label{fig:Observe_rate}
\end{figure}

After completing the initial VRSQ, the first session commences. Observers rate the visual quality of each test image using the five-point Absolute Category Rating (ACR) scale defined in BT.500 recommendation~\cite{ITU-R_BT500_15}, where 5 denotes an excellent quality and 1 a bad quality. Each participant evaluates 300 AR stimuli per session. As the stimuli are static images, the average session duration ranged from 30 to 40 minutes depending on individual rating speed. Participants are free to pause or discontinue at any point according to their personal state. The study comprises four sessions, and participants are encouraged to complete no more than one session per day to minimize fatigue. To reduce context and ordering effects, the stimulus sequence is randomly permuted for each observer~\cite{sendjasni_phd}. All sessions are conducted using the same experimental protocol and equipment, ensuring consistent viewing conditions across the entire study. Figure~\ref{fig:Observe_rate} presents stereoscopic views captured from the headset during the observation and rating phases.

\vspace{-0.5cm}
\subsection{Subjective Test Display}
\vspace{-0.1cm}
The subjective experiments are conducted using a Varjo~VR-3 headset, which is composed of a central $1920\times1920$ uOLED focus panel and a $2880\times2720$ LCD peripheral panel for each eye, providing a $115^{\circ}$ field of view. In our visual pipeline, all stimuli are rendered exclusively on the peripheral LCD displays, bypassing the high-resolution focus panel to ensure uniform presentation across observers. The headset offers 93\% DCI-P3 Color-gamut coverage, an automatic interpupillary-distance range of 59–71~mm, and a 90~Hz refresh rate. Its optical design contributes to reducing eye strain and mitigating simulator sickness~\cite{VarjoVR3}. Stimulus playback is supported by a dedicated high-performance workstation equipped with an Intel Core i9-14900KF processor, 32 GB of DDR5 RAM, and an NVIDIA GeForce RTX 4090 GPU, ensuring real-time rendering and stable framerates throughout all sessions. The interactive AR/VR environment and experimental interface are developed using the Unity game engine.

\vspace{-0.5cm}
\subsection{Subjects and Training}
\vspace{-0.1cm}
Thirty-six participants (22 males and 14 females) were recruited with ages ranging from 18 to 41, with the majority between 23 and 28, and peaks at 24 and 25. No participant had prior access to the test images, and none reported regular use of AR or VR headsets. All participants are compensated in accordance with institutional ethics guidelines.

\vspace{-0.5cm}

\section{Data Analysis}
\vspace{-0.1cm}
This section presents the processing and statistical analysis of the subjective ratings collected in ARIQA-3DS. We describe the data-processing pipeline and MOS computation, assess rating consistency using the Standard Deviation of Opinion Scores (SOS) hypothesis, and evaluate discriminability through non-parametric significance testing. Finally, we examine cybersickness progression across sessions to contextualize participant comfort throughout the experiment.

\vspace{-0.5cm}
\subsection{Data Processing}
\vspace{-0.1cm}
After collecting the subjective ratings, we performed outlier detection and subject rejection in accordance with ITU-R~BT.500-15~\cite{ITU-R_BT500_15}. For each image~$j$, we first evaluated the kurtosis~$\beta_{2,j}$ of the raw scores to determine whether the score distribution follows a normal distribution. %The kurtosis is computed as:

If the computed kurtosis $\beta_{2,j}$ lays between two and four, the distribution is considered normal, and any rating outside $\bar{s}_j \pm 2\,\sigma_j$ is flagged as an outlier ($\bar{s}_j$ is the mean score of image~$j$); otherwise, the thresholds are widened to $\bar{s}_j \pm \sqrt{20}\,\sigma_j$. These outlier ratings, which are in total $2.09\%$ of the collected data, are removed before subsequent analysis. Subject reliability is then assessed using the BT.500 $P_i/Q_i$ criterion. If $s_{ij} \geq \bar{s}_j + 2\sigma_j$, then $P_i = P_i + 1$, whereas if $s_{ij} \leq \bar{s}_j - 2\sigma_j$, then $Q_i = Q_i + 1$. A subject is considered unreliable and rejected only if both $\frac{P_i + Q_i}{N} > 0.05$ and  $\left|\frac{P_i - Q_i}{P_i + Q_i}\right| < 0.3$ held, where $N$ is the total number of rated images. In our experiment, none of the participants met these criteria, and thus, no subject was rejected. After outlier removal and confirmation of subject reliability, each subject’s raw scores are normalized using Z-score transformation $z_i = \frac{s_i - \mu_i}{\sigma_i}$ where $\mu_i$ and $\sigma_i$ denote the mean and standard deviation of all ratings provided by subject~$i$. Assuming most values lie within the range $[-3,3]$, the Z-scores are linearly scaled to $[0,100]$ as $z_i' = \frac{100\,(z_i + 3)}{6}$ following~\cite{ariqa_pico}.  Finally, the MOS for each image is obtained by averaging the scaled Z-scores across all subjects $\mathrm{MOS} = \frac{1}{N}\sum_{i=1}^N z_i'$.

\vspace{-0.5cm}
\subsection{MOS Analysis}
\begin{figure}[t]
    \centering
    % Subfigure A: All Images
    \begin{subfigure}{\linewidth}
        \centering
        \includegraphics[width=0.9\linewidth]{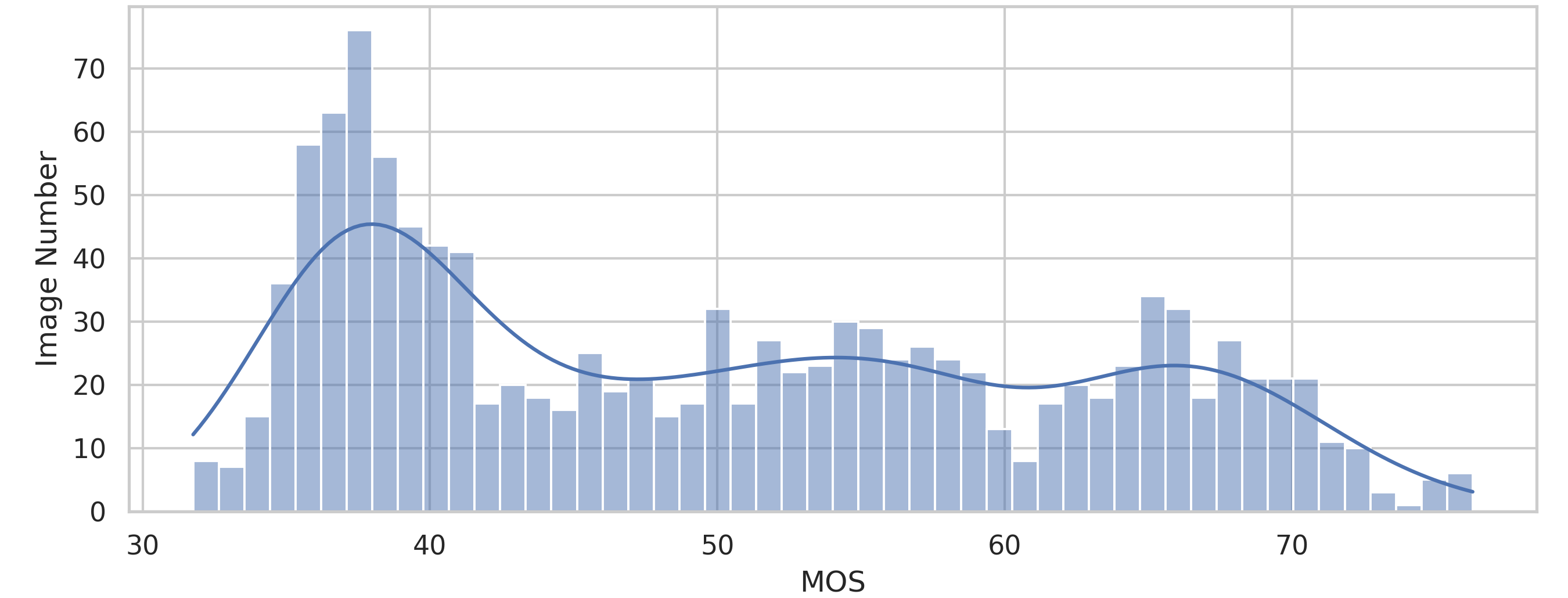}
        \caption{\footnotesize All images ($N = 1200$)}
        \label{fig:mos_dist_all}
    \end{subfigure}

    % Subfigure B: alpha = 0.5
    \begin{subfigure}{\linewidth}
        \centering
        \includegraphics[width=0.9\linewidth]{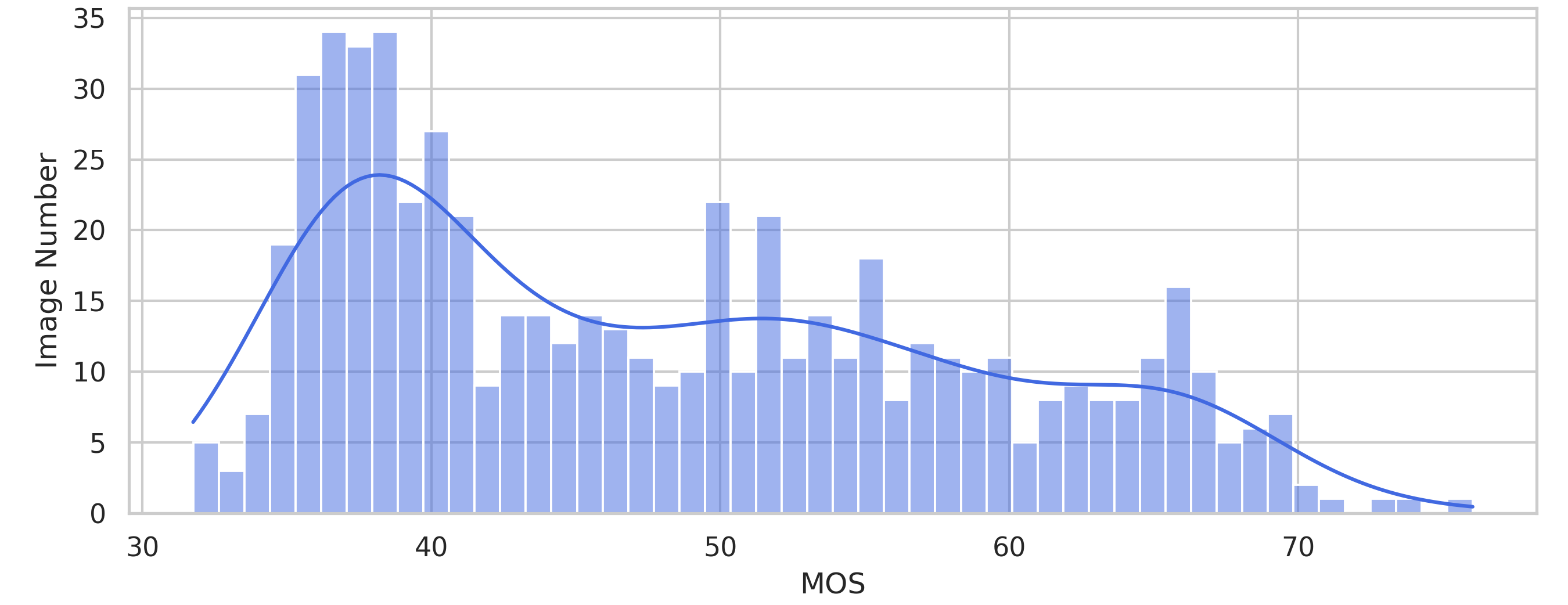}
        \caption{\footnotesize Blending value $\alpha = 0.5$ ($N = 600$)}
        \label{fig:mos_dist_a05}
    \end{subfigure}

    % Subfigure C: alpha = 0.75
    \begin{subfigure}{\linewidth}
        \centering
        \includegraphics[width=0.9\linewidth]{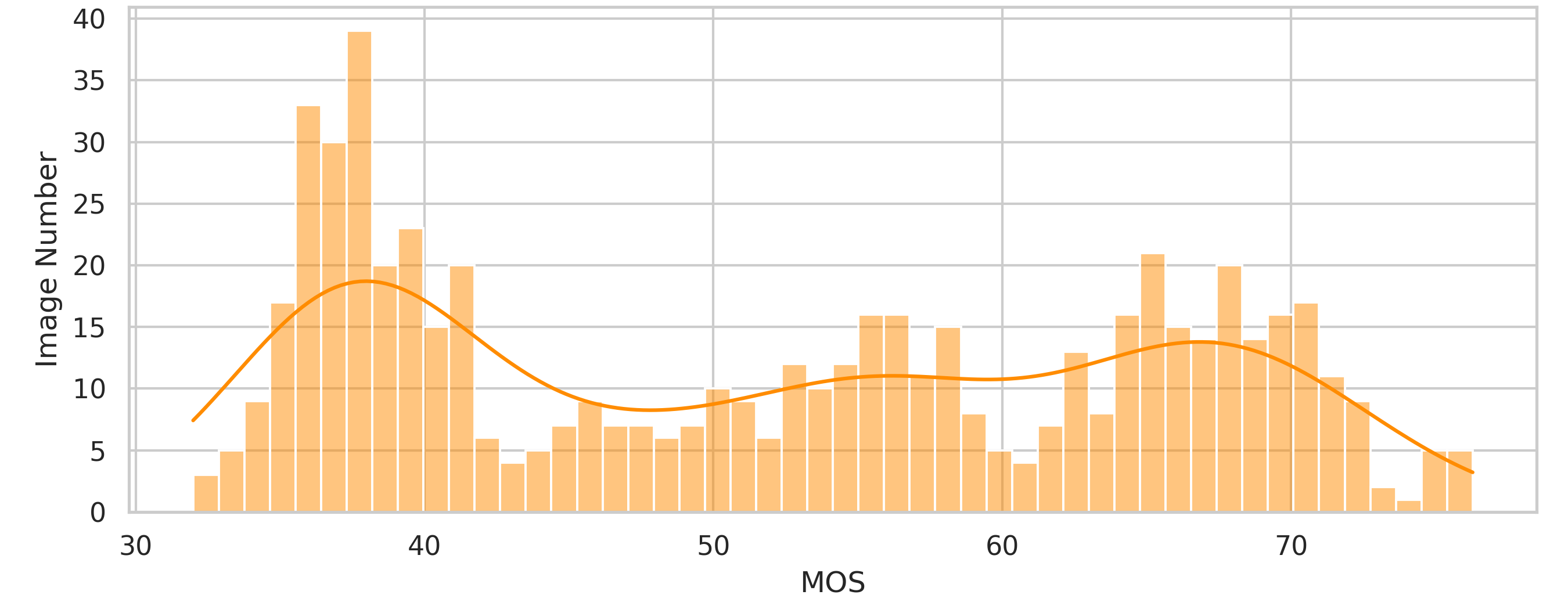}
        \caption{\footnotesize Blending value $\alpha = 0.75$ ($N = 600$)}
        \label{fig:mos_dist_a075}
    \end{subfigure}

    \caption{MOS distributions for (\subref{fig:mos_dist_all}) all images, (\subref{fig:mos_dist_a05}) $\alpha = 0.5$, and (\subref{fig:mos_dist_a075}) $\alpha = 0.75$.\vspace{-0.5cm}}
    \label{fig:mos_dist}
\end{figure}

\vspace{-0.1cm}
\label{sec:mos_analysis}
Figure~\ref{fig:mos_dist} presents the distribution of MOS for all 1200 ARIQA-3DS test images, as well as the separate distributions for the two transparency levels, $\alpha = 0.50$ and $\alpha = 0.75$. The global distribution exhibits a broad, multimodal shape with a prominent cluster around MOS values of 40–45 and a long tail extending toward higher perceived quality. This spread reflects the diversity of AR conditions represented in the dataset, including stereoscopic backgrounds with varying structural and chromatic complexity, multiple foreground categories, and controlled distortions applied to both layers. When analyzing MOS as a function of blending strength, a clear separation emerges between the two transparency levels. At $\alpha = 0.50$, a medium transparency, any degradation in the stereoscopic panorama becomes more prominent. Reduced foreground salience also makes the interaction between layers more ambiguous, increasing the likelihood of binocular visual confusion. As a result, the MOS distribution for $\alpha = 0.50$ is heavily concentrated in the low-to-medium range. Conversely, for $\alpha = 0.75$ the foreground contributes more strongly to the final test image, improving perceptual coherence by providing clearer structural and chromatic cues for the visual system to match across the two eyes. This stronger dominance mitigates competition between layers and reduces fusion instability, producing a broader distribution with many more mid- and high-MOS stimuli. Nevertheless, the persistence of multiple modes indicates that certain scenes remain intrinsically difficult to integrate, especially those involving high-frequency backgrounds or objects affected by severe distortions.

\begin{figure}[t]
    \centering
    \begin{subfigure}[t]{0.4\textwidth}
        \centering
        \includegraphics[width=\linewidth]{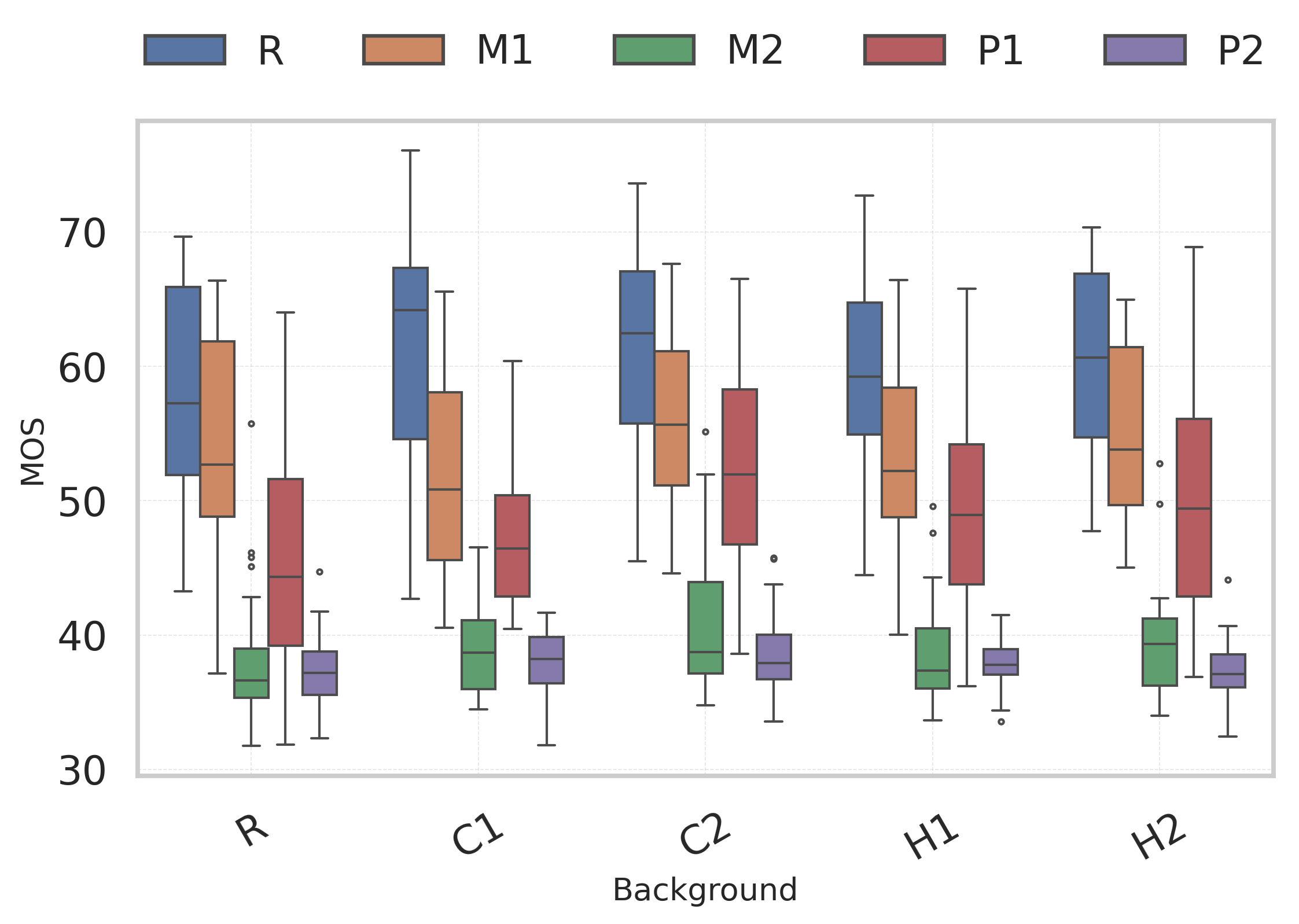}
        \subcaption{$\alpha = 0.5$}
        \label{fig:mos_alpha_05}
    \end{subfigure}
    \vspace{0.5em}
    \begin{subfigure}[t]{0.4\textwidth}
        \centering
        \includegraphics[width=\linewidth]{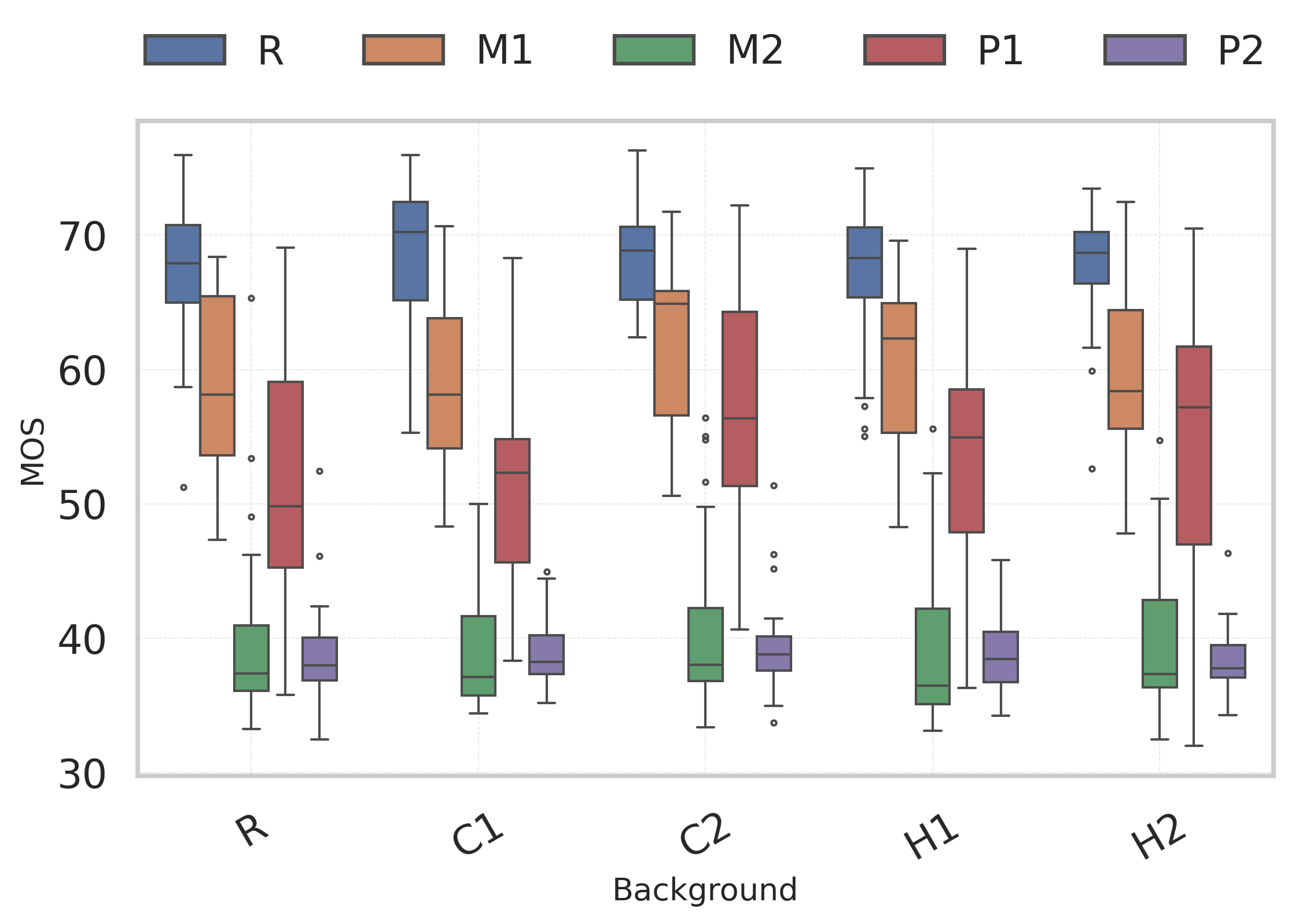}
        \subcaption{$\alpha = 0.75$}
        \label{fig:mos_alpha_075}
    \end{subfigure}
     \caption{Distribution of MOS across background (BG) and foreground (FG) degradation levels. BG degradations include color saturation (C1, C2) and HEVC compression (H1, H2), while FG degradations include motion blur (M1, M2) and pixelation (P1, P2). The reference image is denoted by 'R'. Results are presented for two foreground transparency levels.\vspace{-0.5cm}}
    \label{fig:mos_per_deg}
\end{figure}

To better understand what affects perceived quality, Figure~\ref{fig:mos_per_deg} shows the MOS values for different foreground (FG) and background (BG) distortion levels at both transparency settings. Overall, FG distortions have the strongest impact on quality. Compared to the reference (R), both motion blur levels (M1, M2) and the pixelations (P1, P2) clearly reduce the MOS, with the pixelation yielding the lowest scores. Pixelation strongly changes the object’s structure. For example, text in screenshot elements becomes hard to read, and fine details in graphical objects (such as archaeological models) become unclear, making the object harder to recognize. Motion blur mainly softens edges and removes fine details. The order of severity is stable across all backgrounds: $M2 < M1 < P1 < P2$, meaning that stronger distortions always result in lower MOS. This pattern remains the same for all BG types, showing that the foreground is the main property when judging AR scenes, which aligns with the primary objective of AR applications. Background distortions have a much smaller effect. When the FG is fixed, the differences between the reference background and the color-saturation (C1, C2) or HEVC compression (H1, H2) levels are small, and their MOS ranges often overlap. Color changes only cause minor quality differences, and even strong HEVC compression affects the ratings far less than FG distortions. This suggests that observers mainly looked at the augmented content when giving their scores, and BG distortions alone rarely reduced the perceived quality in a clear way. When the transparency increases from $\alpha = 0.5$ (Figure~\ref{fig:mos_alpha_05}) to $\alpha = 0.75$ (Figure~\ref{fig:mos_alpha_075}), MOS values rise for all conditions. However, this increase does not change the order of distortions or make BG effects stronger. Instead, it simply reflects the higher visibility of the augmented content. Thus, the MOS results show that perceived quality in ARIQA-3DS depends mainly on the foreground. Background distortions only have a small secondary effect, and they become noticeable mostly when the foreground is already degraded.

\vspace{-0.5cm}
\subsection{SOS Analysis}
\vspace{-0.1cm}
The SOS hypothesis~\cite{sos} states that, for properly conducted subjective quality experiments, the variability of observer ratings follows a predictable square-law relationship with respect to MOS. This relationship is governed by a single parameter $a$, which summarizes the overall inter-subject rating variability within a dataset. The SOS hypothesis is defined as:
\begin{equation}
    \mathrm{SOS}(x)^2 = a \left(-x^{2} + 6x - 5\right),
    \label{eq:sos_hypothesis}
\end{equation}
where $x$ denotes the MOS on a five-point ACR scale. The parabolic curve described by Equation~\eqref{eq:sos_hypothesis} reaches its maximum at the midpoint scale ($x=3$), reflecting the empirical observation that rating disagreement is typically highest for stimuli of moderate perceived quality and lowest near the extremes scale, where observer judgments tend to converge. The parameter $a$ controls the height of the curve, with larger values indicating stronger inter-subject variability, greater perceptual ambiguity, or increased task difficulty.

\begin{figure}[t]
    \centering
    \includegraphics[width=0.45\textwidth]{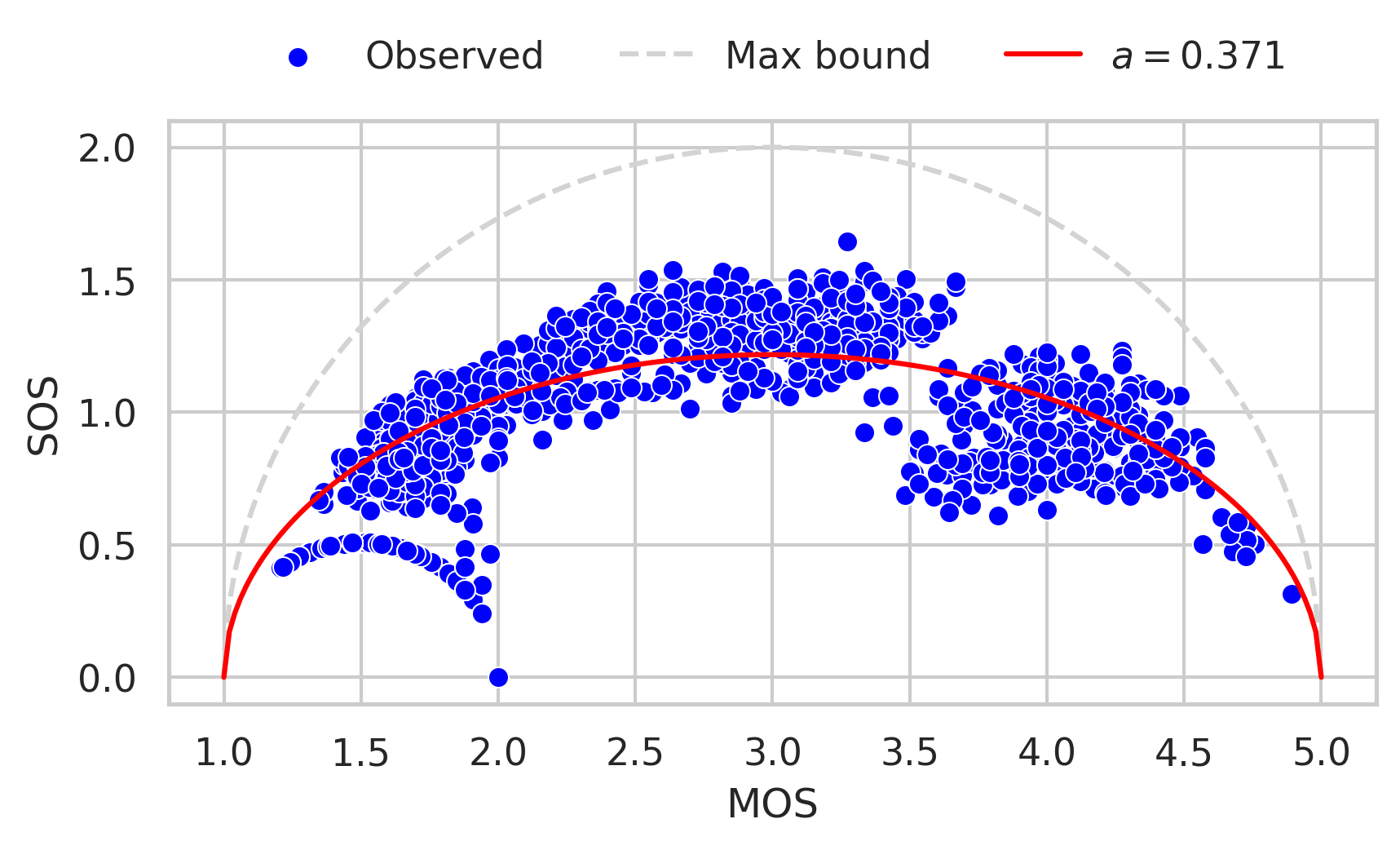}
    \caption{Relationship between MOS and SOS. The fitted curve follows the SOS hypothesis, with parameter $a = 0.371$.\vspace{-0.5cm}}
    \label{fig:sos_analysis}
\end{figure}

Figure~\ref{fig:sos_analysis} illustrates the relationship between MOS and SOS for all test stimuli. The observed data closely follow the parabolic structure predicted by the SOS hypothesis, and fitting the model yields an SOS parameter of $a = 0.371$. This value is substantially higher than those commonly reported for conventional IQA experiments (e.g., $a \in [0.04, 0.17]$) \cite{sos}, indicating significantly increased observer variability in the AR quality assessment task. Such behavior is expected due to the heterogeneous nature of AR scenes, the coexistence of multiple distortion sources, foreground–background blending, visual confusion effects, and the use of head-mounted displays, as well as varying participant familiarity with AR technology. Overall, the strong agreement between the observed SOS distribution and the theoretical model
confirms the inner consistency of the subjective study, while the elevated value of $a$ highlights the increased perceptual complexity inherent to AR image quality evaluation.

\vspace{-0.5cm}
\subsection{Subjective Scores Discriminability Analysis}
\begin{figure}[t]
    \centering
    \includegraphics[width=0.5\textwidth]{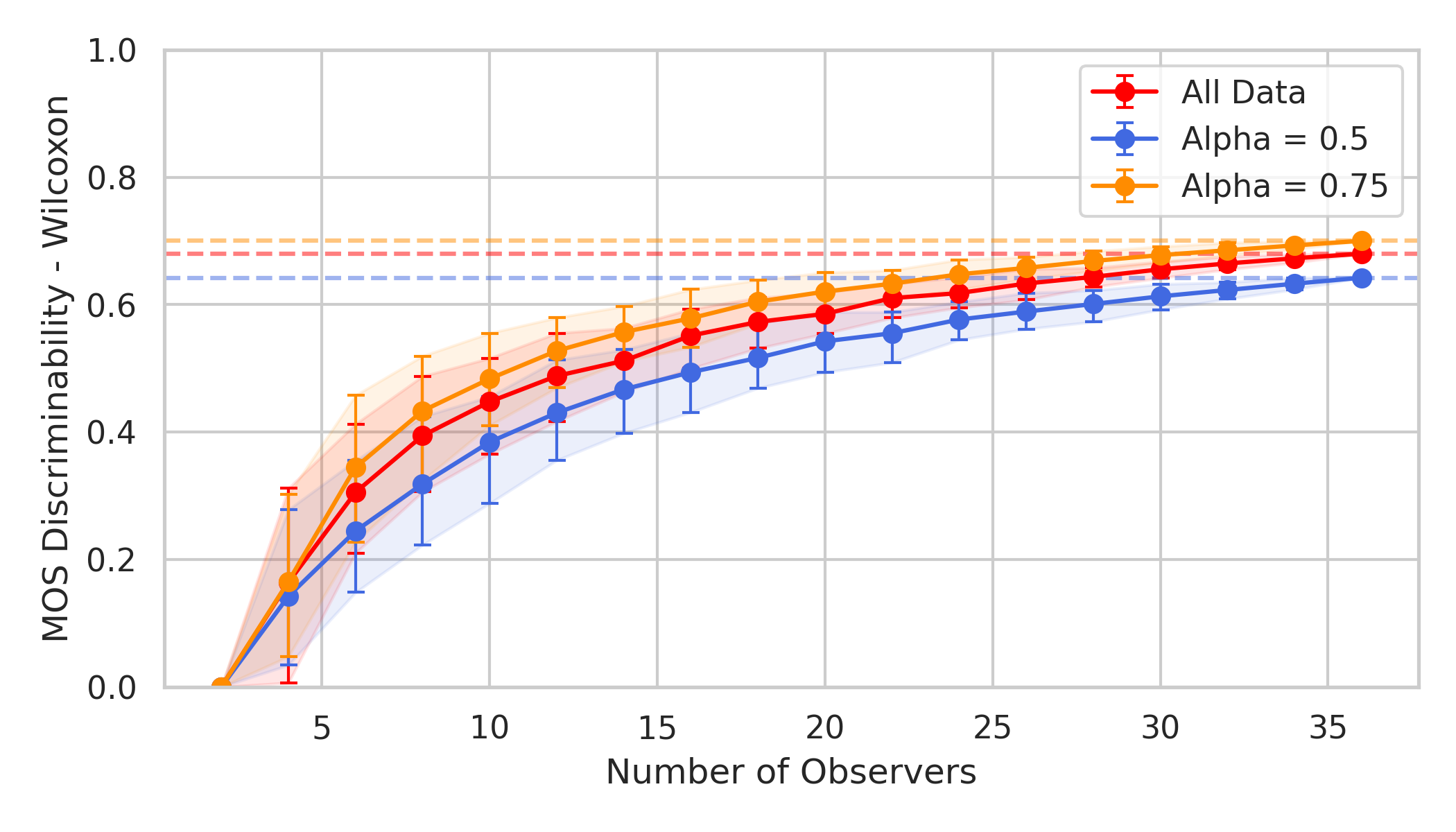}
    \caption{The evolution of the percentage of significantly different pairs (y-axis) with an increasing number of observers (x-axis) for the three experiments. The curves represent mean percentages, and the error bars represent $95\%$ confidence intervals over 100 simulations.\vspace{-0.5cm}}
    \label{fig:disc}
\end{figure}

\vspace{-0.1cm}
The discriminability analysis~\cite{nehme2020comparison} of the ARIQA-3DS dataset quantifies the proportion of stimulus pairs whose Mean Opinion Scores (MOS) differ significantly ($p < 0.05$). Specifically, a two-sample Wilcoxon rank-sum test (also known as the Mann--Whitney $U$ test) is applied to all possible stimulus pairs while progressively increasing the number of observers $N$ from 2 to 36. For each value of $N$, $100$ random observer permutations are generated to ensure statistical robustness, and the average percentage of significantly different stimulus pairs is reported. Figure~\ref{fig:disc} illustrates the evolution of MOS discriminability as a function of the number of observers for the full dataset (\textit{All Data}) as well as for transparency-filtered subsets (\(\alpha = 0.5\) and \(\alpha = 0.75\)). All curves exhibit a characteristic logarithmic growth behavior: discriminability increases rapidly for small observer counts, followed by diminishing gains as $N$ increases. In particular, the rate of improvement slows markedly beyond approximately $25$ observers, with discriminability stabilizing in the range of $65\%$--$70\%$, depending on the transparency condition. Among the subsets, the higher transparency condition (\(\alpha = 0.75\)) consistently achieves slightly higher discriminability, whereas \(\alpha = 0.5\) yields lower values across all observer counts, indicating increased perceptual ambiguity at lower transparency levels. The convergence and stabilization of the curves beyond approximately $25$--$30$ observers indicate that the experimental design is sufficiently powered to produce statistically reliable MOS estimates. This plateau behavior suggests that increasing the number of observers beyond this range yields only marginal improvements in pairwise discriminability, which is consistent with ITU-T recommendations for conducting subjective quality assessment experiments using head-mounted displays (HMDs)~\cite{ITU-P919}.

% \begin{figure*}[t]
%     \centering
%     \includegraphics[width=\textwidth]{subjective/ssq_analysis/vrsq_boxplots_all_sessions.png}
%     \caption{Cybersickness progression measured with VRSQ across session timepoints.}
%     \label{fig:vrsq}
% \end{figure*}
\vspace{-0.5cm}
\subsection{Cybersickness Analysis}
\vspace{-0.1cm}
\label{subsec:ssq}

\begin{table*}[t]
    \centering
    \vspace{-10pt}
    \renewcommand{\arraystretch}{0.9}
    \setlength{\tabcolsep}{4pt}
    \fontsize{9pt}{\baselineskip}\selectfont
    \caption{Mean and median VRSQ scores (O, D, TS) for each session (S1–S4) across the three timepoints (\textit{start}, \textit{middle}, \textit{end}). Higher values indicate greater symptom severity, normalized to the range $[0,100]$. Participants rated each symptom on a five-level scale: None (0), Slight (1), Moderate (2), Strong (3), and Severe (4).}
    \label{tab:vrsq_timepoints}
    % \vspace{-5pt}
    % 2 label columns + (4 + 1 + 4 + 1 + 4) = 16 columns
    \begin{tabular}{cc@{\hskip 15pt}cccccccccccccc}
        \toprule   % oculomotor ($O$), disorientation ($D$) and total score ($TS$),
        Timepoint & Statistic
        & \multicolumn{4}{c}{Oculomotor ($O$)}
        && \multicolumn{4}{c}{Disorientation ($D$)}
        && \multicolumn{4}{c}{Total Score ($TS$)} \\
        \cline{3-6}\cline{8-11}\cline{13-16}
        \rule{0pt}{1.8ex}
        & 
        & S1 & S2 & S3 & S4
        && S1 & S2 & S3 & S4
        && S1 & S2 & S3 & S4 \\
        \midrule

        % ---------------- START ----------------
        \multirow{2}{*}{Start}
        & Mean
            & 10.98 & 8.71 & 8.71 & 7.19  % O Start S1–S4 (mean)
            && 7.87 & 6.06 & 6.96 & 6.06  % D Start S1–S4 (mean)
            && 9.43 & 7.38 & 7.84 & 6.62 \\% TS Start S1–S4 (mean)
        & Median
            & 6.25 & 00.0 & 00.0 & 00.0
            && 5.00 & 00.0 & 00.0 & 5.0
            && 3.12 & 00.0 & 00.0 & 2.5 \\
        \midrule

        % ---------------- MIDDLE ----------------
        \multirow{2}{*}{Middle}
        & Mean
            & 24.62 & 24.05 & 27.08 & 26.70
            && 15.15 & 14.69 & 18.03 & 17.72
            && 19.88 & 19.37 & 22.55 & 22.21 \\
        & Median
            & 18.75 & 18.75 & 18.75 & 25.0
            && 15.00 & 10.0 & 20.0 & 10.0
            && 16.25 & 12.5 & 19.37 & 17.5 \\
        \midrule

        % ---------------- END ----------------
        \multirow{2}{*}{End}
        & Mean
            & 36.93 & 35.03 & 41.28 & 37.5  % O
            && 24.69 & 24.69 & 28.18 & 25.15 % D 
            && 30.81 & 29.86 & 34.73 & 31.32 \\ % TS
        & Median
            & 37.50 & 31.25 & 43.75 & 37.5 % O
            && 25.00 & 20.00 & 20.0 & 20.0 % D
            && 31.25 & 26.25 & 31.87 & 25.62 \\ % TS
        \bottomrule
    \end{tabular}
    % \vspace{-15pt}
\end{table*}

To examine how cybersickness evolved throughout the experiment, we collected VRSQ scores~\cite{SSQ} at three time-points within each session: the \textit{start}, the \textit{middle} (after half of the stimuli were rated), and the \textit{end}. This protocol follows common practice in VR/AR studies, where prolonged exposure can gradually affect user comfort and visual stability. For each time-point, Oculomotor ($O$), Disorientation ($D$), and Total cybersickness ($TS$) scores are computed and normalized to the range $[0,100]$. The resulting mean and median values across the four sessions (S1–S4) are reported in Table~\ref{tab:vrsq_timepoints}.

As shown in Table~\ref{tab:vrsq_timepoints}, oculomotor symptoms ($O$) increased consistently over time in all sessions. Mean $O$ scores are low at the \textit{start} (approximately $7$--$11\%$), rose to the mid-$20\%$ range in the \textit{middle}, and reached their highest values at the \textit{end} (approximately $35$--$41\%$ corresponding to a slight effect). Median scores followed the same upward trend, confirming that this increase is systematic rather than driven by outliers. A repeated-measures ANOVA revealed a significant effect of time-point ($p < 0.001$), indicating that sustained stereoscopic viewing progressively increased visual strain across sessions. Disorientation symptoms ($D$) exhibited a similar pattern. As summarized in Table~\ref{tab:vrsq_timepoints}, mean $D$ scores remained below $8\%$ at the \textit{start}, increased to approx. $15$--$18\%$ by the \textit{middle}, and reached $25$--$28\%$ at the \textit{end}. Median values showed modest inter-session variation but preserved the same monotonic increase over time. The repeated-measures ANOVA confirmed a significant main effect of time-point ($p < 0.001$), reflecting the accumulation of spatial instability and discomfort with repeated exposure to depth-varying AR content. $TS$, aggregating all symptom categories, followed the same progression. According to Table~\ref{tab:vrsq_timepoints}, mean $TS$ values increased from $7$--$9\%$ at the \textit{start}, to $19$--$23\%$ in the \textit{middle}, and to approximately $30$--$35\%$ at the \textit{end}. Median scores closely mirrored this trend across all sessions. Statistical analysis confirmed a significant effect of time-point ($p < 0.001$), indicating a gradual and consistent increase in overall cybersickness throughout the experiment. Overall, Table~\ref{tab:vrsq_timepoints} clearly shows a progression from low discomfort at the beginning of each session to moderate symptoms by the \textit{end}. This temporal evolution is consistent with prior findings on simulator sickness and immersive-display fatigue in VR and AR environments~\cite{Chang20102020, gutierrez2021p919, iana_chaker}. Within the ARIQA-3DS framework, these results indicate that participants completed the study under gradually increasing cybersickness levels, which remained mostly in the slight-to-moderate range and did not compromise the validity of the collected subjective quality ratings.

%The resulting benchmark offers comprehensive ground‑truth quality ratings for stereoscopic AR and provides a much‑needed platform for developing and evaluating new objective metrics.
\vspace{-0.45cm}
%-----------------------------------------------------
\section{Conclusion and Future Work}
In this paper, we introduced ARIQA‑3DS, the first stereoscopic augmented‑reality image quality dataset.  The latter contains 1,200 unique AR viewports produced by compositing 60 diverse foreground objects with 20 omnidirectional background panoramas and systematically varying distortion type, severity, and blending transparency.  We carried out the largest subjective experiments using a video see‑through headset in this area, collecting MOS and VRSQ from 36 participants in accordance with ITU recommendations.  

The analysis of ARIQA‑3DS demonstrates that the subjective ratings are both robust and informative. Outlier rejection removed less than 3\% of scores, and no participants are deemed unreliable. Cybersickness monitoring revealed that oculomotor and disorientation symptoms increased over the course of each session, while remaining within slight-to-moderate discomfort levels at the end of each session. Our perceptual data show that degradations applied to the augmented content exert a stronger influence on perceived quality than those applied to the stereoscopic background, and that higher foreground opacity intensifies visual confusion. Several avenues of future research emerge from this work.  First, ARIQA‑3DS can be extended to dynamic stimuli by incorporating stereoscopic video sequences, head tracking, and temporal distortions to study how motion and attention affect perceived quality over time.  Incorporating multimodal cues such as spatial audio, haptic feedback, and physiological signals like eye‑tracking or heart‑rate variability would enable a more holistic modeling of AR quality of experience. Second, motivated by the success of TransformAR~\cite{sekhri_mmsp} and ARaBIQA~\cite{arabiqa}, future work will focus on evaluating classical and learnable IQA metrics and developing more advanced approaches that account for AR-specific quality features.

%We aim to explore deeper transformer architectures and self-supervised pre-training strategies to further improve the prediction of visual confusion and blending artifacts.

\vspace{-0.4cm}
\section*{Acknowledgments}
This work is partially funded by the Nouvelle-Aquitaine Research Council under project REALISME AAPR2022-2021-17027310. 

%\newpage
\vspace{-0.5cm}
\bibliographystyle{IEEEtran}
\bibliography{ref}

\vspace*{-3\baselineskip}
\begin{IEEEbiography}[{\includegraphics[width=1in,height=1.25in,clip,keepaspectratio]{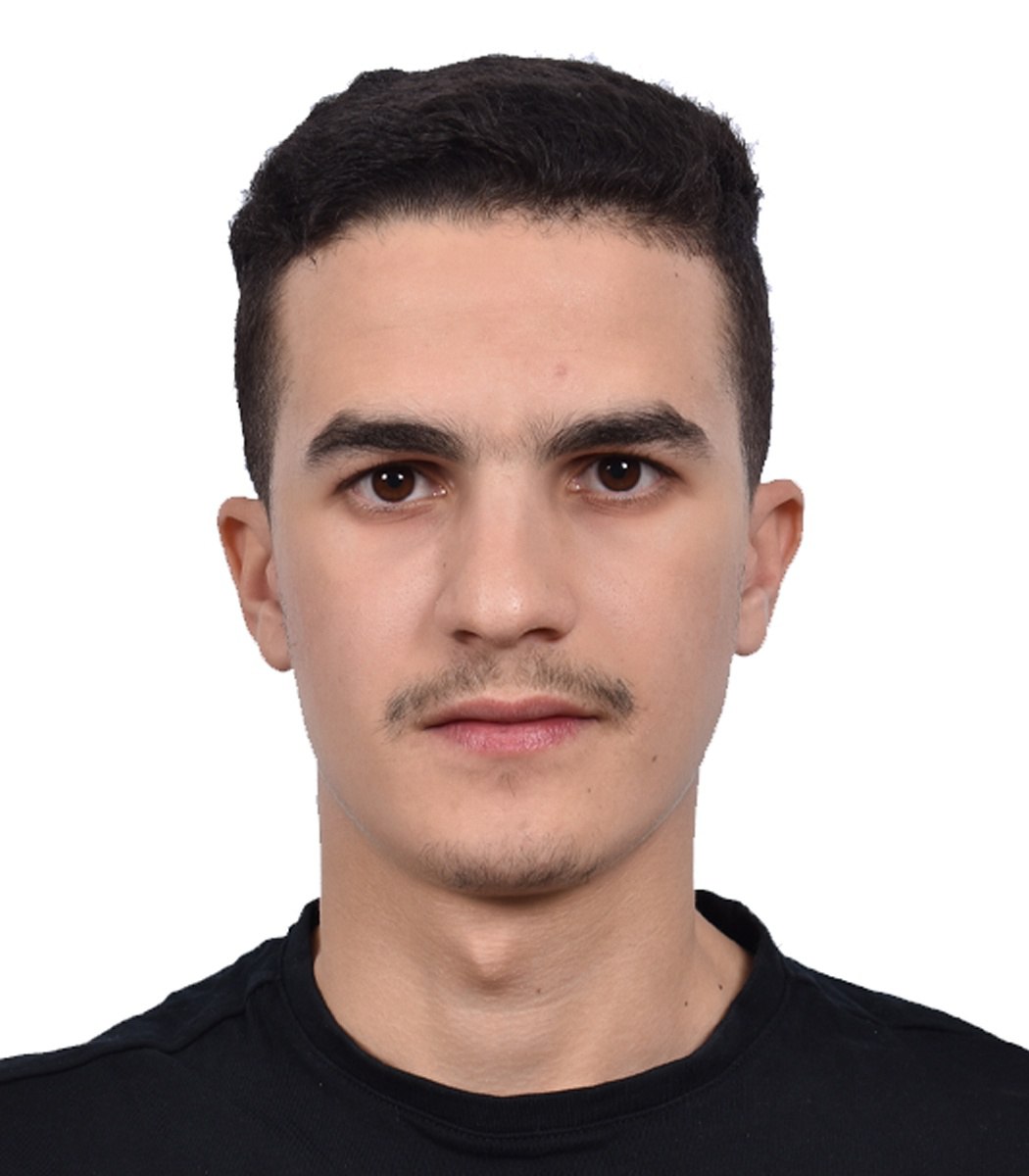}}]{Aymen Sekhri}
was born in Tamalous, Algeria, in 2000. He received his engineering degree from the National Higher School of Telecommunications and ICT (ENSTTIC), Oran, in 2023. He is currently pursuing a Ph.D. degree jointly with the University of Poitiers, France, and the Norwegian University of Science and Technology (NTNU), Norway. His research interests include quality assessment for immersive media, specifically augmented reality, artificial intelligence, and their applications.
\end{IEEEbiography}
\vspace*{-3\baselineskip}
\begin{IEEEbiography}[{\includegraphics[width=1in,height=1.25in,clip,keepaspectratio]{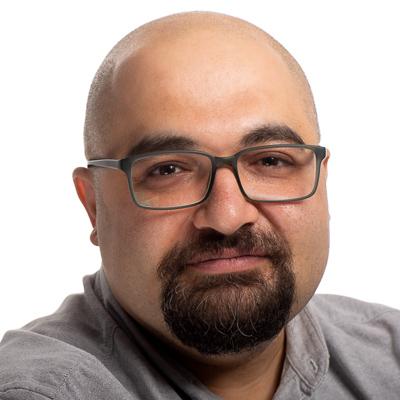}}]{Seyed Ali Amirshahi} 
received his BsC in Electrical Engineering in 2008 from Amirkabir University of Technology, Iran. He graduated from the Master Erasmus Mundus CIMET (Color in Informatics and MEdia Technology) program in 2010. He completed his PhD studies at the Computer Vision Group at the Friedrich-Schiller University of Jena, Germany in 2015. In 2016 He was a postdoctoral fellow at the International Computer Science Institute (ICSI) in Berkeley, California. From 2017 to 2019 he was employed at the Norwegian university of Science and Technology (NTNU) as a FRIPRO/Marie Sklodowska-Curie postdoctoral fellow and a visiting researcher  at the University of Paris 13 (Institut Galil\'ee) Sorbonne Paris Cite. He is currently a Full Professor at  the Department of Computer Science at NTNU in Gjøvik. He is also a member of the Colorlab. His work is focused on image and video quality assessment. 
\end{IEEEbiography}
\vspace*{-3\baselineskip}
\begin{IEEEbiography}[{\includegraphics[width=1in,height=1.25in,clip,keepaspectratio]{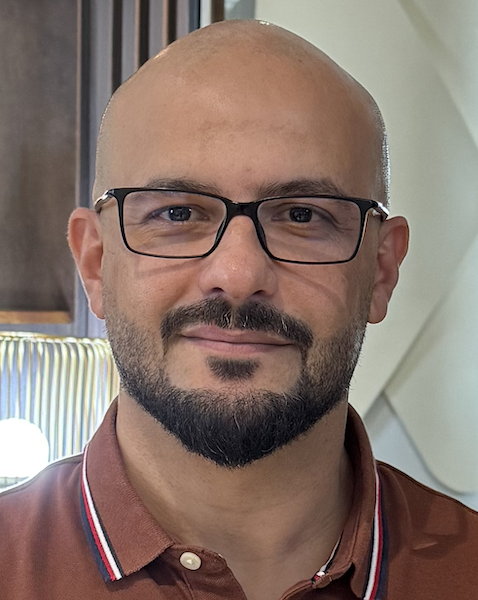}}] {Mohamed-Chaker LARABI} is a Full Professor at Université de Poitiers, in the XLIM CNRS institute, where he is head of the ASALI research axis. He received his PhD from the University of Poitiers in 2002. His scientific interests span different fields of image and video processing, including quality assessment, compression, optimization, and enhancement, traditional and learning-based, for various types of content, including immersive media. 
He supervised or is supervising 20+ PhDs, and he has published over 200 papers. He participated as a PI on various national/international projects. Chaker Larabi is a member of JPEG, MPEG, and AOM. He serves/has served as Associate Editor for IEEE Transactions on Circuits and Systems for Video Technology, IEEE ACCESS, Elsevier Signal Processing: Image Communication, and Elsevier Journal of Visual Communication and Image Representation. He serves currently as Senior Associate Editor for IEEE Trans. on Image Processing.  He regularly serves as area chair and reviewer for several IEEE and EURASIP conferences. Chaker Larabi is an elected member of the MMSP, IVMSP technical committees of the IEEE Signal Processing Society, MSA TC of the IEEE Circuits and Systems Society, as well as the Technical Area Committee on Visual Information Processing of EURASIP. He participated in the organization of several conferences in different roles.
\end{IEEEbiography}
\vfill

\end{document}